\documentclass[prd,aps,amssymb,amsmath,tightenlines,nofootinbib]{revtex4}
\usepackage{graphicx,epsfig}
\usepackage{slashed}
\usepackage{graphicx,epsfig}
\usepackage{slashed}
\usepackage[colorinlistoftodos]{todonotes}
\usepackage{appendix}

\newcommand{\be}{\begin{equation}}
\newcommand{\ee}{\end{equation}}

\begin{document}

\preprint[{\leftline{KCL-PH-TH/2023-{\bf 24}}}

\title{Stringy Running Vacuum Model and current Tensions in Cosmology}

\author{Adri\`a G\'omez-Valent$^{a}$}
\author{N. E. Mavromatos$^{b, c}$}
\author{Joan Sol\`a Peracaula$^d$}

\affiliation{$^a$ INFN, Sezione di Roma 2, and Dipartimento di Fisica, \\Universit\`a di Roma Tor Vergata,
via della Ricerca Scientifica 1, 00133, Roma, Italy,\\
$^b$Physics Department, School of Applied Mathematical and Physical Sciences, National Technical University of Athens, Athens 157 80, Greece,\\
$^c$Theoretical Particle Physics and Cosmology Group, Department of Physics, King's~College~London, Strand, London WC2R 2LS, UK,\\
$^d$Departament de F\'\i sica Qu\`antica i Astrof\'\i sica, \\ and \\ Institute of Cosmos Sciences (ICCUB), Universitat de Barcelona, Av. Diagonal 647 E-08028 Barcelona, Catalonia, Spain.}

\begin{abstract}
 We discuss the potential alleviation of {\it both} the Hubble and the growth of galactic structure data tensions observed 
 in the current epoch of Cosmology in the context of the so-called Stringy Running Vacuum Model (RVM) of Cosmology. 
 This is a gravitational field theory coupled to matter, which, at early eras, contains gravitational (Chern-Simons (CS) type) anomalies and torsion, arising from the fundamental degrees of freedom of the massless gravitational multiplet of an underlying microscopic string theory. The model leads to RVM type inflation without external inflatons, arising from the quartic powers of the Hubble parameter that characterise the vacuum energy density due to primordial-gravitational-wave-induced anomaly CS condensates, and dominate the inflationary era. 
 In modern eras, of relevance to this work, 
 the gravitational anomalies are cancelled by chiral matter, generated at the end of the RVM inflationary era, but cosmic radiation and other matter fields 
 are still responsible for a RVM energy density with terms exhibiting a quadratic-power-of-Hubble-parameter dependence, but also products of the latter with logarithmic $H$-dependencies, arising from potential quantum-gravity and quantum-matter loop effects. In this work, such terms are examined phenomenologically from the point of view of the potential alleviation of the aforementioned current tensions in Cosmology. Using standard information criteria, we find that these tensions can be substantially alleviated in a way consistent not only with the  data, but also with the underlying microscopic theory predictions, associated with the primordial dynamical breaking of supergravity that characterise a pre-RVM-inflationary phase of the model.  
\end{abstract}

\date{\today}

\maketitle

\section{Introduction
\label{sec:1}}
The standard (also called `concordance') model of cosmology, aka $\Lambda$CDM, has been a rather successful  paradigm for the description of the Universe at a pure phenomenological level for more than three decades\,\cite{peebles:1993}, although it became definitively strengthened only in the late nineties\,\cite{Turner:2022gvw}, see particularly \cite{KraussTurner1995,OstrikerSteinhardt1995}.   Despite its phenomenological success, crucial ingredients  such as the presumed existence  of dark matter (DM),  still lack of direct observational evidence. The theoretical situation with the cosmological term $\Lambda$ in Einstein's equations is no less worrisome, though. The main difficulty most likely stems from its interpretation  as a quantity which is connected with the vacuum energy density (VED), $\rho_{\rm vac}$. The proposed connection between the two quantities is well known: $\rho_{\rm vac}=\Lambda/(8\pi G_N)$,  where $G_N$ is Newton's gravitational coupling, usually assumed to be constant. Treating $\Lambda$ as a mere fit parameter one can determine it consistently using different observational sources such as e.g. distant type Ia supernovae (SNIa) and the anisotropies of the cosmic microwave background (CMB) \cite{Riess:1998cb,Perlmutter:1998np,Planck:2015fie,planck}.  Now the VED  is a fundamental concept in  Quantum Field Theory (QFT) and the lack of proper understanding of its connection with cosmology is at the root of the longstanding Cosmological Constant Problem\,\cite{Weinberg:1988cp,Peebles:2002gy,Sola:2013gha}, a problem which is actually not solved with any alternative form of dark energy\,\cite{Solalatest}.

However, new theoretical approaches to some of the above conundrums might be helpful to redefine the difficulties and maybe to alleviate some of these problems. For example, we wish to focus here on the Running Vacuum Model (RVM)~\cite{Sola2,Solalatest,fossil,Sola:2011qr,Sola:2013gha,Sola:2014tta,Sola:2015rra,Sola:2016zeg}, an approach which has been providing  a consistent cosmological framework for a long time (for the latest developments the reader is referred to  \cite{Solalatest}). The RVM could be conceived as an effective field theory description of a smooth evolution of the Universe~\cite{lima1,lima2,Sola:2013gha} from an inflationary epoch, without invoking \textit{ad hoc} inflaton fields, up to the modern era, where the model leads to observable, in principle, deviations from the $\Lambda$CDM paradigm~\cite{Sola:2016jky,rvmpheno,rvmpheno2} (see also \cite{rvmpheno3,rvmpheno4,Geng:2017apd,Tsiapi} for fits of general dynamical dark energy cosmologies to the data). The RVM framework is also capable of 
explaining the entropy production and, in general, thermodynamical properties of the Universe as being consequences of the decay of the running vacuum \cite{lima1,therm1,therm2,therm3}. More recently, the RVM acquired a more fundamental status, which goes beyond the above phenomenological description, in that the VED advocated in it can be derived within the context of renormalizable quantum field theory in curved spacetime, including the appropriate terms (in fact, emerging as quantum effects derived from the effective QFT action) that can trigger inflation in a dynamical way in the early Universe, as shown in \cite{Moreno-Pulido:2020anb,Moreno-Pulido:2022phq,Moreno-Pulido:2022upl,Moreno-Pulido:2023ryo} -- see also \cite{Solalatest} for the essentials.
Remarkably, the RVM can arguably provide
viable alternatives to the $\Lambda$CDM at late epoch, with in-principle observable deviations from it, compatible with the current phenomenology, as well as Big-Bang-Nucleosynthesis (BBN) data~\cite{BBN}, which remarkably lead to similar order of magnitude of the RVM parameters with those of current-era phenomenology. The RVM framework also provides potential resolutions to the recently observed, persisting tensions in the current-epoch cosmological data~\cite{tens1,tens2,tens3}, provided the latter do not admit mundane astrophysical and/or statistical explanations~\cite{mund}. In particular, some versions of the RVM with a low-redshift interaction between the vacuum and cold dark matter or a mildly time-varying effective gravitational ``coupling'' with a renormalized value at cosmological scales, can provide a resolution of these tensions \cite{rvmtens,rvmtens2}.


As pointed out above, the RVM as a successful phenomenological framework can be derived within the context of  QFT in  Friedmann-Lema\^itre-Robertson-Walker (FLRW) spacetime~\cite{Moreno-Pulido:2020anb,Moreno-Pulido:2022phq,Moreno-Pulido:2022upl,Moreno-Pulido:2023ryo}.  It would, however, be desirable to derive the various coefficients parametrising the RVM energy density
also from alternative microscopic descriptions. One interesting possibility is the embedding of the RVM into microscopic string-inspired gravitational models with axions and gravitational anomalies in the early Universe~\cite{bms1,bms2,ms1,ms2} (for reviews and comparison 
with other approaches to quantum gravity and string theory~\cite{string}, specifically theories with torsion~\cite{torsion,kaloper}, see \cite{MLV,Mphil,Mtorsion}). Both approaches, QFT in curved spacetime and string-inspired gravitational theories can be simultaneously applicable, in the sense that matter QFT corrections to the gravitational effective actions, arising from integrating out massive quantum fields, can also be implemented in  string effective field theory constructions. There is a crucial difference between these two approaches, though. In the latter, the so-called Stringy Running Vacuum Model (StRVM), the dominant {terms in the cosmic vacuum energy density during the early Universe evolve with $H^4$ due to quantum-gravity-induced anomaly condensates. In contrast, in the local QFT approach, gravity is considered as a background, and it is matter which is viewed as quantum and integrated over in a path-integral framework. In the QFT approach, in the context of scalar field theories with non-minimal coupling to space-time curvature, one obtains explicit $H^6$ and higher powers in the vacuum energy density \cite{Moreno-Pulido:2022phq,Moreno-Pulido:2023ryo}. So far, 
the explicit $H^4$ RVM corrections to the vacuum energy density have been reproduced explicitly only within the StRVM framework, as well as a one-loop supergravity framework in a de Sitter background~\cite{rvmsugra,tseytlin,houston1,houston2}.
\footnote{We note for completion that, outside the RVM framework, $H^4$ terms in cosmology can arise~\cite{GUP1,GUP2} through quadratic curvature corrections in the effective actions induced in certain scenarios of the generalised uncertainty principle that may characterise quantum-gravity models, {\it e.g.} the quantum-gravity proposal of K.~Stelle~\cite{stelle}.} The latter will be of crucial importance in this work, as it will provide a prototype theory, in which, on integrating out gravitational degrees of freedom, we obtain (in approximately de Sitter cosmological backgrounds corresponding to a slowly varying Hubble parameter $H(t)$) terms in the effective cosmic energy density of the form $H^2\, {\rm ln}(H)$. Moreover, such supergravities may characterise pre-RVM inflationary eras in the StRVM framework, playing an important r\^ole in providing microscopic origin of the chiral metric fluctuations leading to the aforementioned anomaly condensates~\cite{ms2}.   

As mentioned previously, from a microscopic point of view, explicit examples of RVM have been given in the context of the so-called StRVM~\cite{bms1,bms2,ms1,ms2} parity-violating string-inspired cosmologies of Chern-Simons type~\cite{jackiw,yunes}, which are characterised by the presence of chiral mixed (gravitational and gauge) anomaly terms, as a result of the Green-Schwarz-mechanism~\cite{gs} for the cancellation of anomalies~\cite{alvarez}
in string theory. The theory also contains torsion, which is dual to the gravitational (also known as a string-model-independent) axion field~\cite{kaloper,svrcek}. 

In this context, it is the condensate of anomalous gravitational Chern-Simons terms, induced by chiral, parity-violating gravitational waves~\cite{stephon,lyth} that induce non-linear terms of fourth order $H^4$ in the Hubble parameter of a cosmological background in the cosmic energy density.\footnote{This would also be true in the presence of populations of rotating black holes~\cite{YunBH1,YunBH2,chatz,chatz2}, and in general
space-time deformations which are ``chiral'' in nature.} Such $H^4$ terms, which are dominant at early epochs, drives RVM inflation without the need for external inflaton fields~\cite{lima1,lima2}. In \cite{bms1,ms1,MLV} we have discussed the possibility of forming a condensate of the gravitational anomaly, in case there is a {\it macroscopic number} of sources of gravitational waves (GW), with constructive interference. In \cite{bms1,ms1} an $\mathcal O(1)$ proper density of defects ({\it i.e.} over the proper cosmological volume) was assumed, which lead for consistency to large (compared to Planck scale) string mass scales $M_s \sim 10^{-3}\, M_{\rm Pl}$.
In \cite{MLV}, on the other hand, we have assumed proper macroscopic densities of sources of such GW in the expanding Universe, $n_\star \equiv  \frac{\mathcal N(t)}{\sqrt{-g}}$,   which implies a total number of sources $N= \int d^4x \, \mathcal N (t) = \int d^4 x \sqrt{-g} \, n_\star$. As discussed in that work, in such a case, the string scale $M_s$ is a free parameter, to be fixed phenomenologically, and can also serve as the Ultra-Violet (UV) cutoff of the string-inspired effective point-like field theory.

It is the point of this work to analyse the current-epoch phenomenology of StRVM, which, notably, in addition to integer even powers of $H$, also contains 
logarithmic dependent terms $H^2 {\rm ln}(H^2)$~\cite{ms1,Mtorsion}. In the StRVM such terms arise exclusively from quantum-gravity corrections. A prototype quantum field theory model where such corrections have been computed explicitly is the one-loop supergravity~\cite{tseytlin,houston1,houston2}, which, as already mentioned, yields a RVM theory when placed in approximately de Sitter cosmological backgrounds with slowly varying Hubble parameter~\cite{rvmsugra}. Apart from serving as prototypes, where such quantum-graviton corrections can be computed reliably, such supergravities play an important r\^ole also for the physics of StRVM, providing a microscopic origin of the primordial chiral gravitational waves that lead to the Chern-Simons condensates which drive the RVM inflationary period~\cite{ms1,ms2}. 
We stress that such logarithmic corrections $H^2\,{\rm ln}(H)$ also appear in QFT by integrating out matter quantum fields~\cite{Moreno-Pulido:2020anb,Moreno-Pulido:2022phq,Moreno-Pulido:2022upl,Moreno-Pulido:2023ryo}. In the current work we constrain the StRVM by making use of an updated and rich set of cosmological measurements, and study the potential alleviation of the Hubble tension and the tension with the large-scale structure data. We shall see that the dynamical scale of supergravity breaking (which occurs during the pre-RVM-inflation era of the model~\cite{ms1}) can be constrained this way to lie close to the reduced Planck scale, exactly as expected in the microscopic underlying model. We find this a remarkable fact, worthy of stressing to the reader.

The structure of the article is as follows: in the next section \ref{sec:preinfl} we discuss the one-loop supergravity model in de Sitter backgrounds, as a provider of a prototype model of calculable quantum-gravity induced one-loop corrections to the one-loop effective action. In section \ref{sec:rvmmodern} we explain how this approach can lead to non-polynomial, logarithmic,  corrections of the form $H^2\,{\rm ln}(H)$ to the late-eras cosmic vacuum energy density, when the supergravity model is embedded in a cosmological context. We then proceed in section \ref{sec:strvm} to apply the above considerations to the (one-loop) quantum-gravity corrected effective cosmological field theory of the StRVM, discussing in detail the formalism underlying the late-era equation of state (EoS) and the related phenomenology, including the linearly perturbed gravitational equations. In particular we carefully discuss the perturbation formalism that will allow us to use the Einstein-Boltzmann code \texttt{CLASS} \cite{CLASS1,CLASS2} so as to perform fits of the StRVM to the available cosmological data. As a follow up, we discuss the potential alleviation of the present-era cosmological tensions. In section \ref{sec:DataMethod} we describe the datasets and the methodology employed to constrain the model. In section \ref{sec:results} we present our results. In section \ref{sec:ectcc}, we discuss the status of the energy conditions in the StRVM, and also argue how the model might evade the so-called {\it transplanckian cosmic censorship conjecture} (TCC). Finally section \ref{sec:concl} contains our conclusions and outlook. Technical aspects of our study are given in two Appendices.  

\section{One-loop corrected Supergravities in pre-RVM Inflationary eras as a prototype model for discussing quantum-gravity corrections\label{sec:preinfl}} 

As discussed in \cite{ms1,ms2} there is a pre-RVM-inflationary epoch in the StRVM, which may be characterised by dynamically broken local supersymmetry (supergravity (SUGRA)), through, {\it e.g.}, condensates of the gravitino field $\psi_\mu$, the supersymmetry partner of graviton. 
Such extensions are still within the spirit of \cite{bms1} that only fields from the gravitational multiplet of strings appear as external fields in effective field theories at the very early stages of the string-inspired Universe.  An important aspect of the dynamically broken supergravity model is that quantum-gravity corrections, {\it i.e.} corrections arising from integrating out massless spin-2 graviton fields, in the one-loop effective action of SUGRA considered in a de Sitter background~\cite{tseytlin,houston1,houston2} lead to the appearance of terms in the effective action exhibiting logarithmic dependence on the cosmological constant. As discussed in \cite{houston1,ms2}, such SUGRA de Sitter vacua are {\it metastable}, characterised by calculable (within the field theory approximation)  imaginary parts, which eventually imply tunnelling of the system into the RVM inflationary 
anomaly-condensate-induced RVM inflationary spacetime. This is in agreement with the no-go theorems on the incompatibility of stable de Sitter space-time vacua with SUGRA models arising from compactifications in string/brane models~\cite{nunez}.

The one-loop effective potential of a prototype N=1 d=4-dimensional SUGRA model, which is dynamically broken by means of a condensate 
of gravitinos $\psi_\mu$:
\begin{align}\label{gravinocond}
\sigma_c = \kappa \, \langle \overline \psi_\mu \, \psi^\mu \rangle ~,
\end{align}
is given, in a Euclidean (E) path-integral formalism, in a background spacetime with a one-loop-renormalised (positive) cosmological constant $\Lambda > 0$, by~\cite{houston1,houston2}:
\begin{align}\label{effactionl3}
\Gamma^{\rm (E)} \simeq&-\frac{1}{2\kappa^2} \int d^4 x \sqrt{\widehat g_{\rm E}}
\left[\left(\widehat R-2\Lambda_1 \right)  +\alpha_1 \, \widehat R+
\alpha_2 \, \widehat R^2\right]~,
\end{align}
where $\kappa=1/M_{\rm Pl}$ is the inverse of the reduced Planck mass, and we used the fact that the curvature
scalar in the (Euclidean) four-dimensional de Sitter spacetime is given by: 
\begin{align}\label{rl}
\widehat R = 4\Lambda,
\end{align}
or:
\begin{align}\label{cosdS}
\widehat R = 12 \bar{H}^2, \quad \bar{H} = {\rm constant}
\end{align}
in the cosmological de Sitter case, with a constant Hubble parameter $\bar{H}$, of interest to us here, and 
in a specific gauge. 

The remaining quantities in \eqref{effactionl3} are given by ~\cite{houston2} 
    \begin{align} \label{LL1}
\Lambda_{1}=-\, \kappa^2 \,
\left(-\frac{\Lambda_0}{\kappa^2}+\alpha_0^F+\alpha_0^B\right)~,
    \end{align}
where the superscripts $B$ anf $F$ refer to terms arising from integration of massless (quantum) gravitons  and gravitinos, which are Bosonic (B) and Fermionic (F) degrees of freedom, respectively. Formula \eqref{LL1} expresses one-loop corrections to the bare cosmological constant $\Lambda_0$,   
\begin{align}\label{l0tree}
	 \frac{\Lambda_0}{\kappa^2} 	
	 \equiv \sigma_c^2  -f^2 ~,
\end{align}
with
\begin{align}\label{a0}
		\alpha_0^{F} &= {\kappa}^4 \, \sigma_c^4 \, \Big(0.100\,  \ln \left( \frac{{\kappa}^2 \, \sigma_c^2}{3 \mu_\tau^2}\right) + 0.126 \Big)~, \nonumber \\
		 				\alpha_0^{B} &= \kappa ^4 \, \left(f^2-\sigma_c^2\right)^2 \left(0.027 - 0.018 \ln \left(\frac{3 \kappa ^2 \left(f^2-\sigma_c^2\right)}{2 \mu_\tau^2}\right)\right)~,
		\end{align}
		and
\begin{align}\label{alphasugra}
         \alpha_1=\frac{\kappa^2}{2}\left(\alpha^F_1+\alpha^B_1\right)~,\quad
        \alpha_2=\frac{\kappa^2}{8}\left(\alpha^F_2+\alpha^B_2\right)~,
    \end{align}
    where
\begin{eqnarray}\label{aif}
\alpha^F_1&=& 0.067\, \kappa^2 \sigma_c ^2  -0.021\,
\kappa^2 \sigma_c ^2 \, {\rm ln}
\left(\frac{\Lambda}{\mu_\tau^2}\right)
 +  0.073\, \kappa^2 \sigma_c ^2 \, {\rm ln}
\left(\frac{\kappa^2\sigma_c^2}{\mu_\tau^2} \right)~,
\nonumber \\
\alpha^F_{2}&=& 0.029 + 0.014\, {\rm ln} \left(\frac{\kappa^2\sigma_c^2}{\mu_\tau^2}\right) 
-0.029\, {\rm ln} \left(\frac{\Lambda}{\mu_\tau^2}\right)~, \nonumber \\
\alpha^B_1&=& -0.083 \Lambda_0 + 0.018\, \Lambda_0 \, {\rm ln} \left(\frac{\Lambda }{3 
\mu_\tau^2}\right)  + 
0.049\, \Lambda_0\,  {\rm ln} \left(-\frac{3 \Lambda_0}{\mu_\tau^2}\right)~, \nonumber \\ \alpha^B_{2} &=& 0.020 +  0.021\, {\rm ln} \left(\frac{\Lambda }{3 \mu_\tau^2}\right) - 
0.014\, {\rm ln} \left(-\frac{6 \Lambda_0}{\mu_\tau^2}\right)~.
\end{eqnarray}
We note that the tree-level (bare) cosmological constant $\Lambda_0$ \eqref{l0tree}
must be necessarily \emph{negative}, given that (unbroken) supergravity (local supersymmetry), which characterises the classical (Euclidean) action $S^{(E)}_{\rm cl} = -\frac{1}{2\kappa^2} \int d^4 x \, \sqrt{\widehat g_E}\, \Big(\widehat{R} - 2\Lambda_0\Big) $, is incompatible with de Sitter vacua~\cite{houston1,houston2,tseytlin}. The one-loop renormalised cosmological constant 
$\Lambda$, on the other hand, is {\it positive}, due to quantum corrections (see Eq.~\eqref{LL1}), and this is compatible with the case of dynamically-broken supergravity~\cite{wittendbs}. The quantity $\sigma_c < f$, where $f$ is the energy scale of dynamical breaking of supergravity (and also global supersymmetry)~\cite{houston2}, denotes the value of the gravitino condensate field \eqref{gravinocond} at the minimum of its double-well one-loop effective  potential.

The replacement of $\Lambda$ in all the above expressions by the scalar curvature, \eqref{rl}, is understood. The quantity $\mu_\tau^2$ (with dimensions of mass-squared) is an inverse Renormalization Group (RG)  scale, in the sense that it is a  UV ultraviolet cutoff on the proper-time $\tau$, which regularises UV divergences~\cite{houston1,houston2,tseytlin}, that is, small $\mu_\tau^2$ values correspond to the UV regime of the theory, whilst large $\mu_\tau^2$ values correspond to the infrared (IR).\footnote{The alert reader should note that in this article, for economy in  notation, we use the same symbol $\tau$ also to denote the conformal time in the expanding-universe metric \eqref{confmetr}, used in section \ref{sec:perturbations} and in Appendix \ref{sec:appendixPert}.} In other words, as we 
flow from UV to IR the value of $\mu_r$ increases, which is the opposite behavior of an ordinary RG scale in QFT.
In the work of \cite{houston1,houston2}, supergravity breaks dynamically at a large scale $\mu_\tau^2$ close to Planck scale~\cite{houston1}, which allows us to set  from now on
\begin{align}\label{muPl}
\mu_\tau^2 \sim M^2_{\rm Pl} = \kappa^{-2}.
\end{align}
In the dynamically-broken-supergravity phase, the gravitino and its condensate acquire large masses, which can be well above the grand-unification scale, even close to Planck scale for our purposes~\cite{ms1,ms2,rvmsugra}, since we want supergravity to be broken well before the RVM inflation. 
This  can be arranged for appropriate values of the scale $f$.

From a physical viewpoint, the dynamically broken supergravity phase of the Universe during the pre-RVM inflationary era, entails a 
(not necessarily slow-roll) first, hill-top, inflation~\cite{Ellis}, near the origin ($\sigma_c=0$) of the gravitino double-well potential~\cite{ms1}. The latter
is responsible for the homogeneity and isotropy of the Universe assumed for the RVM inflation, which is induced by condensation of primordial GW.
These GW are due to either merging of primordial black holes, or the non-shperically-symmetric collapse (or collisions) of domain walls, that in turn arise in scenarios~\cite{ms1} in which the initial degeneracy of the two vacua of the gravitino 
condensate double-well potential is lifted, {\it e.g.} due to cosmic percolation effects~\cite{perc1,perc2} (see also \cite{perc0,perc0b}). 

As discussed in detail in \cite{ms2}, quantum fluctuations of the condensate field $\sigma_c$, which exceed $f$, lead to imaginary parts in the effective action \eqref{effactionl3} (due to negative arguments of the 
appropriate logarithms in \eqref{aif}), which in turn imply that the spontaneously broken SUGRA de Sitter vacuum is {\it metastable}, 
leading to a {\it tunnelling} of the system to the RVM inflation, induced by the GW condensation of the gravitational anomaly terms.  
The latter is itself metastable, due to the time dependence of the Hubble parameter, leading eventually to the exit from the second RVM inflationary phase, as a consequence of the decay of the RVM vacua. Such metastable vacua are in agreement with the rigorous no-go theorems concerning SUGRA and de Sitter vacua~\cite{nunez}, or the swampland conjectures~\cite{swamp1,swamp2,swamp3,swamp4,swamp5}, concerning embedding of RVM into UV complete theories of quantum gravity, such as string theory, which are thus avoided in the case of the RVM. In this way one may avoid the cosmological constant problem. In this respect, we also mention that similar conclusions can be reached in the QFT approach to RVM in the context of theories involving matter fields in curved (cosmological) spacetimes, whereby the novel renormalization studies of \cite{Moreno-Pulido:2020anb,Moreno-Pulido:2022phq,Moreno-Pulido:2022upl,Moreno-Pulido:2023ryo} also lead to the avoidance of the appearance of a cosmological constant term in the renormalised effective action, in contrast to conventional approaches. 

The conjectural (at this stage) interpretation that the $\ln\Lambda$ terms in the effective action of the dynamically-broken supergravity theory are  not simple coefficients of curvature terms dependent on the renormalised $\Lambda$, but can themselves be viewed as covariant curvature scalars also away from a de Sitter background \eqref{rl}, implies that the quantum supergravity effective action \eqref{effactionl3} now has non-polynomial terms of the form
\begin{align}\label{logterms}
\widehat R^{n}\, {\rm ln}(\kappa^2 \widehat R)\,, \, n=1,2\,, 
\end{align}
on account of \eqref{muPl}.\footnote{The reader should notice that our modified gravity has a Minkowski flat limit, as the curvature scalar $R \to 0$, and in this respect it has to be contrasted with the purely ${\rm ln}R$ gravity suggested in \cite{nojiri}, whose terms grow with small curvatures.}  

Moreover, in the broken supergravity phase, which also includes the RVM inflation, the gravitino and its condensate fields, are superheavy excitations in our stringy scenario~\cite{ms1,ms2,rvmsugra}, with masses close to the Planck mass, and this can be integrated out in the path integral. 
This in turn will imply Planck-mass suppression for these terms in the action of the effective field theory, leaving only terms of the massless degrees of freedom, eventually leading to the string-inspired action of \cite{bms1,ms1,ms2} with axion and graviton degrees of freedom (the dilaton has been assumed stabilised to a constant, in a self consistent way~\cite{bms2}).

\section{Implications for Modern Eras \label{sec:rvmmodern}}

It is important to stress that, as one observes from \eqref{alphasugra} and \eqref{aif}, terms with logarithmic dependence on the Hubble parameter $H$, \eqref{logterms}, are also the result of integrating out massless graviton fluctuations. Thus, we may also encounter such modified relativity (bosonic) effective actions in the modern era, where the gravitino fields play no r\^ole~\cite{ms1,Mtorsion}, purely due to quantum gravity (QG) effects.\footnote{Such terms 
also exist in the RVM inflationary era, but they are suppressed compared to the inflation-driving anomaly condensates, scaling like $H^4$~\cite{ms1}. 
Indeed, the QG-induced $H^4 {\rm ln}(\kappa^2 H^2)$ terms in the early Universe are subdominant 
compared to the GW-induced $H^4$ terms in the Chern-Simons condensate for 
$\kappa^4 |\mathcal E_0| < 1$, as required by the {\it transplanckian conjecture}, which the bare scale $\mathcal E_0$ is assumed to satisfy. Thus our conclusions on RVM inflation~\cite{bms1,ms1,ms2} remain unaffected.} 
 Therefore, the result of integrating out massless graviton fluctuations in 
the effective action, which  describes the gravitational dynamics of the post-inflationary stringy RVM Universe, until the current era, leads to weak QG corrections, given by adding to the standard Einstein-Hilbert Lagrangian term one-loop corrections of the form (we analytically continue back to a Minkowski-signature spacetime from now on):\footnote{In this work we follow the convention
for the signature of the metric $(+, -, -, -)$, and the definitions of the Riemann Curvature tensor
$R^\lambda_{\,\,\,\,\mu \nu \sigma} = \partial_\nu \, \Gamma^\lambda_{\,\,\mu\sigma} + \Gamma^\rho_{\,\, \mu\sigma} \, \Gamma^\lambda_{\,\, \rho\nu} - (\nu \leftrightarrow \sigma)$, the Ricci tensor $R_{\mu\nu} = R^\lambda_{\,\,\,\,\mu \lambda \nu}$, and the Ricci scalar $R = R_{\mu\nu}g^{\mu\nu}$. Overall, these
correspond to the $(-, +, +)$ conventions in the popular classification by Misner {\it et al.} \cite{Misner}.}
\begin{equation}\label{1looplagr}
\delta \mathcal L^{\rm 1-loop}_{\rm quant.~grav.} =  - \sqrt{- g}\, \Big[ \widetilde \alpha_0 + R \Big(\tilde c_1 + \tilde c_2\, {\rm ln}\left(-\frac{1}{12}\kappa^2  R\right)\Big) \Big] + \dots 
\end{equation}
where the constant $\widetilde \alpha_0$ plays the r\^ole of a one-loop induced vacuum energy density. 

From the supergravity example~\cite{houston1,houston2} we have seen that $\widetilde \alpha_0 >0$, and 
that the constant coefficients $\tilde c_i$ assume  the form ({\it cf.} \eqref{aif})
\begin{align}\label{cs}
 \tilde c_i \propto \kappa^2 \mathcal E_0, \,\, {\rm or} \,\, \tilde c_i \propto \kappa^2 \mathcal E_0 \, {\rm ln}(\kappa^4 |\mathcal E_0|)\,, \,  i=1,2, 
\end{align}
with $\mathcal E_0$ a bare (constant) vacuum energy density scale.   From \eqref{cs} it becomes clear that the sign of the coefficients $\tilde c_i, i=1,2$ depends on the signature of the bare cosmological constant term. In supergravity models $\mathcal E_0 < 0$, but in the absence of supersymmetry $\mathcal E_0$ could be positive. This will play an important r\^ole for the generic parametrization of our phenomenological analysis of the StRVM at late epochs in section \ref{sec:strvm}. 

The ellipses $\dots $ in \eqref{1looplagr} denote terms of quadratic and higher order in $ R$, which are subdominant in the current epoch $(H = H_0)$, when the Universe enters again a de Sitter phase, but with a much smaller (approximately) de Sitter Hubble parameter. We stress once more, 
that the structures \eqref{1looplagr} appear generic for weak QG corrections about de Sitter backgrounds~\cite{tseytlin}, as appropriate for the current era of the Universe. We may therefore conjecture that the corrections \eqref{1looplagr} can lead to a modified version of the stringy RVM discussed so far, thus playing a r\^ole in the current-era phenomenology.

Indeed, considering the graviton equations stemming from the one-loop corrected effective Lagrangian, we easily observe that the correction terms \eqref{1looplagr} 
imply corrections to the effective stress-energy tensor in the current era of the form: 
\begin{equation}\label{1loopenden}
\delta \rho_0^{\rm vac}  =  \frac{1}{2}\widetilde \alpha_0 + 3 (\tilde c_1-\tilde c_2) \, H_0^2 + 3 \tilde c_2 \, H_0^2 \, \ln{(\kappa^2\,H_0^2}) + \dots~, 
\end{equation}
where the $\dots$ denote subleading terms proportional to $(\dot H_0)^2, \ddot H_0$, which are negligible in the current epoch, during which the Universe enters once again a de-Sitter phase. 

Crucial to us in  this work will be the 
approximately-de-Sitter nature of the spacetime at late eras of the Universe,
which implies a mild but non-trivial 
cosmic-time dependence of the Hubble parameter. Thus in what follows, we shall replace $H_0$ by $H$.
It is also important to notice that the supergravity prototype~\cite{houston1,houston2,rvmsugra} indicates that the one-loop correction (dark-energy-type) term $\frac{1}{2} \widetilde \alpha_0$ is constant, {\it independent} of ${\rm ln} H^2$ terms.

The total stress energy tensor is obtained by adding \eqref{1loopenden} to the stress tensor of the RVM, including a cosmological constant $c_0 > 0$ phenomenologically,\footnote{An important formal remark is in order here.  A cosmological constant $c_0$ is added phenomenologically in \eqref{logH2}, given that in our string-inspired StRVM a de Sitter spacetime is not welcome (perturbative string theory scattering matrix is not well-defined in de Sitter spacetimes, and non-perturbative string vacua are incompatible with de Sitter spacetimes, due to swampland~\cite{swamp1,swamp2,swamp3,swamp4,swamp5}). During the RVM phase there was no $c_0$ term and any (approximate) de Sitter contribution arose dynamically and it was metastable.
Thus, it is most plausible, that this will be the case also of the current era, during which, as a result of the depletion of matter content in the Universe,
gravitational anomalies (which, in the StRVM framework, where cancelled~\cite{bms1,bms2,ms1,ms2} at the exit of inflation from  chiral matter that was created as a consequence of the decay of the RVM vacuum) could resurface. In such a case a de Sitter contribution could also be viewed due to some sort of condensate of GW, which are much weaker of course in the current epoch. At present we do not have a concrete scenario for the current era, and this should be considered as a speculative remark.}
thus obtaining a total energy density (we use below $\kappa^{-2} = M_{\rm Pl}^2$):
\begin{eqnarray}\label{logH2}
  \rho_{\rm RVM}^{\rm vac}\equiv \rho_{DE}   \simeq  3 M_{\rm Pl}^2 \Big\{c_0 + [ \nu + 
d_1\, \ln{(M_{\rm Pl}^2/H^2}) ] \, H^2   + \dots\Big\} ~,  
 \end{eqnarray}
where $c_0 > 0, \nu > 0$, and  $d_1$ are phenomenological parameters. Below we treat these parameters as functions of the cosmic time, because we assume that they can take in principle different values in different epochs of the expansion. This characterises the stringy RVM model, as we have discussed explicitly in \cite{bms1,ms1}, where, for instance, it was shown that the parameter $\nu < 0$ during the RVM inflation, 
as a consequence of the gravitational anomaly contributions, while, par contrast, $\nu > 0$ in the post-inflationary epoch. 
Below, we also argue 
that $\nu$ remains of the same order from the BBN until the modern epochs, in a way consistent with the available cosmological data, provided one assumes the running vacuum to be the dominant source of the currently observed dark energy in the Universe.

Indeed, compatibility of the model with the BBN constraints leads to constraints on $d_1$
in narrow windows~\cite{BBN}:
\begin{align}\label{bbnconst}
d_{1}^{\rm BBN}  \in\left(-1.0\times 10^{-5},1.3\times 
10^{-5} \right)\,,
\end{align}
with $\frac{c_0}{H_{0}^2}\in\left(0.697,0.704\right)$, as required by the 
current-era dark energy constraints~\cite{planck}.\footnote{We remind the reader that during the modern eras, including that of BBN, 
the coefficients of terms of order $H^4$ (and higher) in \eqref{logH2} cannot be constrained. Indeed, as shown in \cite{BBN}, if one assumes that the $H^4$ terms in the energy density \eqref{logH2}, which also include logarithmic corrections due to one-loop quantum-graviton contributions~\cite{houston1}, $\frac{\alpha}{H_I^2} [ 1 + d_2\, 
\ln{(M_{\rm Pl}^2/H^2})] \, H^4,$
play an equal r\^ole to the rest of the terms in \eqref{logH2}, as far as the observed dark energy today is concerned, 
then one finds $\frac{\alpha^{\rm BBN}}{H_I^2} = 10^{46}\, {\rm GeV}^{-2}$, and only then BBN constraints 
imply  $d_{2}^{\rm BBN}  \in\left(-8.5\times 10^{-2},1.2 \times 
10^{-2}
\right)$. Such a value for the coefficient $\alpha_{\rm BBN}$  is far greater than the value such a quantity would have in the RVM inflationary phase ($\alpha/H_I^2 \sim 10^{-26}~{\rm GeV}^{-2}$ for $\alpha \sim \mathcal O(1)$). Hence, only the $d_1$ coefficient can be constrained by BBN data.}
 
Therefore, on ignoring terms of order $\mathcal O(H^4)$ and higher, the current-era vacuum energy density, including QG logarithmic-$H$ corrections,  assumes the form:
\begin{align}\label{logLRVM0}
  \rho_{\rm RVM}^{\rm vac}\equiv \rho_{DE}   \simeq  3 M_{\rm Pl}^2 \Big\{c_0 + [ \nu_0 + 
(d_{1})_0\, \ln{(M_{\rm Pl}^2/H_0^2}) ] \, H_0^2   + \dots\Big\} ~,  
 \end{align}
where $c_0 > 0$, $\nu_0 > 0$ and  $(d_1)_0$,  are phenomenological parameters. 
where, in a standard notation in cosmology, a subscript ``$0$'' indicates present-day quantities. In the context of StRVM, such a form is derived~\cite{bms1} by assuming cosmic radiation fields as playing a r\^ole in inducing the $H^2$ terms. It should be stressed, that in contrast to the stringy RVM inflationary scenario, where the anomaly terms imply a negative $\nu < 0$, the post-inflationary 
$\nu_0$, is {\it positive}. 

We remark at this stage that fitting \eqref{logLRVM0}, in the case $(d_1)_0=0$, to the plethora of the cosmological data
leads to the conclusion that~\cite{rvmpheno,rvmpheno2,rvmtens,rvmtens2}: 
\begin{align}\label{nuc0}
\nu_0 = \mathcal O(10^{-4}-10^{-3}) > 0~, \qquad   3\kappa^2 \, c_0 = \mathcal O(10^{-122})  > 0~.
\end{align}
Remarkably, these order-of-magnitude estimates are consistent with BBN~\cite{BBN}, which indicate that for the phenomenologically correct RVM models,
the coefficient $\nu > 0$ does not change in order of magnitude since, at least, the BBN era.

The presence of the $H_0^2$ terms in \eqref{logLRVM0}
leads to observable in principle deviations from $\Lambda$CDM, since there is  
different scaling of the Hubble parameter today compared to the prediction of the $\Lambda$CDM paradigm:
\begin{align}\label{modern}
H^{\rm modern} (a) = H_0 \Big(  \Big[ 1 - \frac{c_0}{H_0^2\, (1-\nu)}  \Big] \, a^{-3\, (1-\nu)}  + \frac{c_0}{H_0^2\, (1-\nu)} \Big)^{1/2} \equiv H_0 \Big( \widetilde \Omega_{m\, 0} \, a^{-3(1-\nu)} + \widetilde \Omega_{\Lambda\, 0}\Big)^{1/2},
\end{align} 
where the quantities $\widetilde \Omega_{\Lambda\ 0} \equiv \frac{c_0}{H_0^2\, (1-\nu)} >0$ and 
$\widetilde \Omega_{m0} \equiv  1 - \widetilde \Omega_{\Lambda\, 0}$ play the r\^ole of the matter and cosmological-constant energy densities today, in units of the 
critical density of the Universe, with the term $\widetilde \Omega_{\Lambda\, 0}$ {\it dominant} in the {\it current era}.

We also remark at this point that the existence of logarithmic terms, induced by quantum-graviton corrections, in the vacuum energy density ({\it cf.} \eqref{1loopenden}) may lead to some differences with respect to other RVMs whose phenomenology has been recently analyzed in \cite{rvmtens,rvmtens2}. However, this subject requires a more careful consideration in the future, as the QFT version of the RVM contains also logarithmic terms, see the detailed works \cite{Moreno-Pulido:2020anb,Moreno-Pulido:2022phq,Moreno-Pulido:2022upl,Moreno-Pulido:2023ryo}. The phenomenology of the QG-modified stringy RVM, as far as the current-epoch tensions in the cosmological data are concerned, will be the subject of the next section. 

Before moving onto the data analysis, we make a final but important comment, regarding the presence of $H^2 \, {\rm ln}(H^2)$ 
terms in effective running vacuum model energy densities in the context of QFT non-minimally coupled to FLRW spacetime, during the current cosmological epoch, discussed in 
\cite{Moreno-Pulido:2020anb,Moreno-Pulido:2022phq,Moreno-Pulido:2022upl,Moreno-Pulido:2023ryo}. Indeed, as shown in those works, following a RG analysis well within the spirit of the RVM~\cite{Sola:2013gha}, in which $H$ is viewed as a RG scaling parameter, one may write for the RVM energy density $\rho_{\rm RVM}^{\rm QFT}$ in such QFT, connecting two values of the Hubble parameter, an $H$ era, and the current era, at which $H=H_0$:\footnote{Although we show explicitly below the case of a scalar quantum field, non-minimally coupled to gravity~\cite{Moreno-Pulido:2020anb,Moreno-Pulido:2022phq,Moreno-Pulido:2022upl}, qualitatively similar results, as far as the induced logarithmically dependent corrections on the Hubble parameter $H(t)$ are concerned, also characterise fermionic quantum field theories~\cite{Moreno-Pulido:2023ryo}. }
\begin{align}\label{qft}
\rho_{\rm RVM}^{\rm vac~QFT}  (H) = \rho_{\rm RVM}^0 + \frac{3\, \nu_{\rm eff} (H) }{\kappa^2} \, \Big(H^2 - H_0^2 \Big)\,, \quad 
\nu_{\rm eff} (H) \simeq \frac{1}{16\pi^2} \Big(\xi - \frac{1}{6}\Big) \, {m^2}\kappa^2\,  \, {\rm ln}\Big(\frac{m^2}{H^2}\Big)  \,,
\end{align}
 where  $\nu_{\rm eff}(H)$ is not a constant coefficient, but depends mildly on the cosmic time today through the time dependence of $H(t)$.
In the formula \eqref{qft}, 
$\xi$ is the non-minimal coupling of the (quantized) scalar matter fields with gravity, with the conformal theory corresponding to $\xi=1/6$. The quantity $\rho_{\rm RVM}^0 = 
\frac{3}{\kappa^2}(c_0 + \nu_0  \, H_0^2)$ denotes the standard current-era RVM energy density~\cite{rvmpheno,rvmpheno2}.
In arriving at \eqref{qft}, we have assumed ${\rm ln}(m^2/H^2) \gg 1$. 
We mention for completion that the exact formula for scalars can be found in  \cite{Moreno-Pulido:2022phq} and was complemented with the (quantized) fermionic contributions in \cite{Moreno-Pulido:2023ryo}.

We stress that the form \eqref{qft}, is due 
to the appropriate subtractions of the vacuum energy at modern eras~\cite{Moreno-Pulido:2020anb,Moreno-Pulido:2022phq,Moreno-Pulido:2022upl,Moreno-Pulido:2023ryo}. Thus, we observe that the logarithmic corrections to the vacuum energy density due to quantum matter in the RVM framework are of the form $(H^2(t)-H_0^2)\, {\rm ln}(m^{2}\, H(t)^{-2}) \ll H(t)^2\, {\rm ln}(m^{2}\, H(t)^{-2})$, as $H(t) \to H_0$.
Quantitatively, in the modern eras, of relevance to the observed tensions in the data, which correspond to small redshifts $z \ll 1$, in which one has matter and vacuum energy dominance, with equations of state $w_m =0, w_{\rm vac} \simeq -1$, the quantum-matter 
corrections 
in \eqref{qft} may be expressed as:
\begin{align}\label{nueffz}
\delta \rho_{\rm RVM}^{\rm vac \, QFT} 
\simeq \frac{9 m^2}{16\pi^2} \, \Big(\xi - \frac{1}{6}\Big)\, H_0^2 \, z\,  \Omega^{(0)}_m \, {\rm ln}\Big(\frac{m^2}{H^2}\Big) + \dots\,,\quad 0 < z \ll 1 
\end{align}
where the $\dots$ indicate terms of higher order in the redshift $z$. 
In arriving at \eqref{nueffz} we have taken into account the standard relation 
\begin{align}\label{HzH0}
H^2(z) = H^2_0 \Big(\sum_i \Omega^{(0)}_i \, (1 + z)^{3(1 + w_i)}\Big)\,,
\end{align}
where $\Omega_i^{(0)}$, 
$i=m,\Lambda$, are the current-era densities of matter ($m$) and vacuum energy ($\Lambda$) in units of the critical density (the radiation component is assumed negligible today), for which we have $\Sigma_{i=m,\Lambda}\Omega_i^{(0)}=1$, and $w_m=0 \, (w_\Lambda \simeq -1)$ are the corresponding equations of state.

For completion, we also remark that, on using \eqref{cosdS} for an approximately de Sitter cosmological background, we can express \eqref{nueffz} 
in a covariant form in terms of the scalar curvature of spacetime, as:
\begin{align}\label{nueffz2}
 \delta \rho_{\rm RVM}^{\rm vac \, QFT} 
\simeq \frac{m^2}{64 \pi^2}\Big(\xi - \frac{1}{6}\Big)\, (R_0 - R)\, {\rm ln}\Big(\frac{-12m^2}{R}\Big)\,,  
\end{align}
where $R_0 = -12 H_0^2$ is the scalar curvature of the expanding-universe spacetime today. 

From \eqref{nueffz} (or \eqref{nueffz2}) we thus observe that, despite the fact that for ordinary matter $m^2\gg H_0^2\sim R_0 $, the smallness of the prefactor $H^2-H_0^2 \ll H^2_0$ during modern eras (when $z\ll 1$), implies that, depending on the scale $\sqrt{|\mathcal E_0|}$ of the quantum gravity corrections ({\it cf.} \eqref{cs}), the latter may actually be dominant over matter QFT fields, even in modern epochs.  Basically this depends on whether $\kappa^2 \mathcal E_0$ is bigger or not than $m^2(\xi-1/6)$.  To satisfy that dominance it would enforce  $\mathcal{E}_0^{1/4}$ to be in the geometric mean of $M_{\rm pl}$ and $m$. Since $m$ is a GUT scale, it means that  $\mathcal{E}_0^{1/4}$ should be of order of the string scale, $\sim 10^{17}$ GeV (for large string scales, that we work with in the model of \cite{bms1,ms1}). In the context of the StRVM this is to be expected, given that the SUGRA dynamical breaking phase in the stringy RVM preceded the RVM inflation~\cite{ms1}, which should occur around the GUT scale, so that the model is in agreement with the data~\cite{planck}. 
As we shall see in section \ref{sec:perturbations} ({\it cf.} \eqref{eps7}, \eqref{f7}), this range of $\kappa^2 \mathcal E_0$ emerges from requiring that the SUGRA-StRVM model 
alleviates both, the Hubble and the growth-of-structure
tensions. On the other hand,
for generic quantum gravity models, both quantum-gravity and matter-QFT effects could be of comparable magnitude. In the present study, and for the sake of simplicity,  we will consider that the quantum-gravity effects are dominant, as this will help to highlight their potentiality. A full analysis by taking into account both the QG effects and the combined QFT contributions from quantized boson and fermion fields\,\cite{Moreno-Pulido:2023ryo} would be too cumbersome at this stage and can be left for a future study. Let us recall at this point that these QFT effects from matter can also help to relieve the existing tensions in the $\Lambda$CDM context\,\cite{rvmtens,rvmtens2}, so we cannot exclude a collaborative effect from QG and QFT to highly alleviate both tensions.

Although our phenomenological analysis in the following sections will be general as far as the quantum-graviton corections are concerned, nonetheless one can make the plausible assumption~\cite{ms1,rvmsugra,Mtorsion} that the 
primordial supergravity breaking, occurring at a high energy scale $\sqrt{|f|}$ close to Planck scale~\cite{houston1}, actually constitutes the dominant source of the logarithmic (on the Hubble scale $H(t)$) corrections to the effective action for the entire evolution of the StRVM Universe, from the SUGRA-breaking phase till the 
current-era. In such a case, we may identify  
\begin{align}\label{e0f}
|\mathcal E_0 |^{1/4}= \sqrt{|f|}\,,
\end{align}
in \eqref{cs}, and attempt to match with the current cosmological data. In what follows, we shall see that, upon using \eqref{e0f}, such fits yield $\sqrt{f}$ sufficiently close to Planck scale, which is consistent with our dynamical supergravity breaking scenario during the pre-RVM inflationary era in the stringy RVM. Of course, in generic quantum gravity models in de Sitter backgrounds with bare cosmological constant $\kappa^2|\mathcal E_0|$  of order of today's measured cosmological constant, the quantum matter field corrections \eqref{nueffz} may dominate, in which case the relevant phenomenology should be properly adjusted to take account of this fact.

 Before closing the section, an important remark should be made concerning \eqref{qft}. 
This contribution pertains strictly to quantum field theory matter effects, which 
have been calculated by integrating out quantum matter-field fluctuations, with the 
appropriate vacuum subtractions absorbed in $\rho_{\rm RVM}^0$~\cite{Moreno-Pulido:2020anb,Moreno-Pulido:2022phq,Moreno-Pulido:2022upl,Moreno-Pulido:2023ryo}. On the other hand, as discussed in detail in \cite{bms1}, background matter and cosmic radiation fields, can lead to modern eras to contributions proportional to $H^2$ (without the ${\rm ln}(H^2)$ factors  in the vacuum energy density). This important feature will be taken into account in our phenomenological analysis below.

\section{Detailed Late-eras Phenomenology of the Stringy RVM (StRVM) \label{sec:strvm}}

Motivated from the above discussion, in both supergravity and renormalised quantum field theory in curved spacetime, we parameterised both quantum graviton and matter field effects using the following action:\footnote{For convenience, we have expressed the argument of the logarithm in \eqref{eq:action} in terms of the present-era curvature scalar $R_0$, instead of units of the square of the Planck mass scale, $\kappa^{-2}$, which would be the natural scale in a quantum gravity scenario, such as the dynamically-broken supergravity model~\cite{houston1,houston2}, discussed in section \ref{sec:preinfl} ({\it cf.}  \eqref{effactionl3}, \eqref{muPl}). This implies the vanishing of the logarithmic terms in the modern era, and will be taken into account when we match the predictions of the supergravity Lagrangian for the coefficients $c_i$, $i=1,2$, see {\it e.g.} \eqref{c12} below.}
\begin{equation}\label{eq:action}
S=-\int d^4x\,\sqrt{-g}\,\left[c_0+R\left(c_1+c_2\ln\left(\frac{R}{R_0}\right)\right)\right]+S_m\,,
\end{equation}
where, in connection with our previous notation and discussion, $c _1 = \frac{1}{2\kappa^2} + \tilde c_1$ denotes both the tree-level and one-loop-induced quantum-gravity  and background matter QFT corrections to the gravitational constant, whilst $c_2 $ is the corresponding one-loop logarithmic corrections, which are only due to quantum gravity, given that the corresponding matter QFT corrections will be assumed subdominant for the supergravity model, which we restrict our attention to in this work, according to our previous discussion. The constant $c_0$, might depend on the renormalization-group scale, but is independent of $H$, as follows explicitly from the study of the supergravity case. The reader is reminded that, in this and the following sections, 
quantum-matter-field induced loop corrections are assumed subdominant, as compared to their quantum-graviton counterparts.\footnote{If the matter terms were not negligible, as it may happen in generic quantum gravity models in (approximately) de Sitter backgrounds, with bare cosmological constants of order of the observed one today~\cite{planck}, 
then their effects are captured by adding to \eqref{eq:action} a term with the structure:
\begin{align}
 \delta S^{\rm matter~QFT}   = 
 - \int d^4 x \sqrt{-g} \, h_1 \, 
 {\rm ln}\Big(\frac{R}{R_0}\Big)\,,
\end{align}
where $h_1$ is a constant coefficient that is determined from \eqref{nueffz2}. Appropriate additions (which are suppressed by factors $m^2/m_{\rm Pl}^2$) should also be made to the coefficients $c_i, i=0,1,2,$ of \eqref{eq:action} to take account of all the matter contributions, as follows from \eqref{nueffz2}. Of course, the reader should always bear in mind that the matter terms vanish today ($R \to R_0$, $z \to 0$), and thus the constraints on the matter contributions have to be imposed by looking at $z \ne 0$, provided the matter terms there surpass the quantum-graviton contributions from the bare cosmological constant terms. We shall not consider this more general case here, thereby ignoring the matter effects in front of the quantum-graviton-induced effects from primordial supergravity breaking, which we assume to be the dominant effects in our stringy RVM scenario~\cite{ms1,rvmsugra,Mtorsion}.}

The modified Einstein equations, obtained from the variation of the action \eqref{eq:action} with respect to the metric, read,

\begin{equation}
G_{\mu\nu}-\frac{c_0}{2c_1}g_{\mu\nu}+\frac{c_2}{c_1}\left[G_{\mu\nu}\ln\left(\frac{R}{R_0}\right)+R_{\mu\nu}-\nabla_\mu\left(\frac{\partial_\nu R}{R}\right)+g_{\mu\nu}\nabla^\alpha\left(\frac{\partial_\alpha R}{R}\right)\right]=\frac{T^{(M)}_{\mu\nu}}{2c_1}\,,\label{eq:EinsEq}
\end{equation}
with the superscript $(M)$ denoting the joint contribution of non-relativistic matter fields and radiation. We can define the tensor

\begin{equation}
A_{\mu\nu}\equiv -G_{\mu\nu}\ln\left(\frac{R}{R_0}\right)-R_{\mu\nu}+\nabla_\mu\left(\frac{\partial_\nu R}{R}\right)-g_{\mu\nu}\nabla^\alpha\left(\frac{\partial_\alpha R}{R}\right)\label{eq:Atensor}
\end{equation}
such that 

\begin{equation}
T^{\rm (vac)}_{\mu\nu}\equiv c_0g_{\mu\nu}+2c_2A_{\mu\nu}\qquad {\rm and}\qquad G_{\mu\nu}=\frac{T^{(M)}_{\mu\nu}+T^{\rm (vac)}_{\mu\nu}}{2c_1}\,.\label{eq:Tvac}
\end{equation}
The model at this point does not know anything about quantum effects from matter. We have assumed at the level of the action that $c_0$, $c_1$ and $c_2$ are constants, so the model reduces, in practice, to an $f(R)$ model. In $f(R)$ models the total energy-momentum tensor (EMT) is covariantly conserved. Actually, it is easy to see that the geometrical quantity \eqref{eq:Atensor} obeys $\nabla^\mu A_{\mu\nu}=0$ and, therefore, $\nabla^\mu T^{\rm (vac)}_{\mu\nu}=0$. Thus, for the model under consideration, with constant $c_i$'s, there is no cross-talk between the vacuum and matter nor a running of the gravitational coupling. 

Now let us study a more general scenario with potential quantum effects from matter, in which we promote all the constants $c_i$ to functions and consider that matter is not conserved. The conservation equations take the following form, 

\begin{equation}
2G^{\mu}_{\nu}\partial_\mu c_1 = \nabla^\mu T^{(M)}_{\mu\nu}+\partial_\nu c_0+2A^{\mu}_{\nu}\partial_\mu c_2\,.
\end{equation}
According to \eqref{eq:Tvac}, $c_1$ can be associated to the effective gravitational coupling,  $c_1=(16\pi G)^{-1}$. It can be a dynamical quantity if we abandon the assumption of the covariant conservation of matter and/or the constancy of $c_0$ and/or $c_2$. For instance, we could promote $c_0$ to a function, assuming a RVM expression, $c_0(R)=\tilde{c}_0+\tilde{c}_1R$, while keeping $c_2=const.$ and $\nabla^\mu T^{(M)}_{\mu\nu}=0$. The dynamics of the vacuum would depend in this case on the two parameters $c_1$ and $\tilde{c}_1$. But actually it is not clear how to build $T^{\rm (vac)}_{\mu\nu}$ from Eq. \eqref{eq:EinsEq}. There is a high degree of ambiguity. Maybe it is natural to re-express Eq. \eqref{eq:EinsEq} as follows, 

\begin{equation}
G_{\mu\nu}\left[1+\frac{c_2}{c_1}+\frac{c_2}{c_1}\ln\left(\frac{R}{R_0}\right)\right]-\frac{c_0}{2c_1}g_{\mu\nu}+\frac{c_2}{c_1}\left[\frac{R}{2}g_{\mu\nu}-\nabla_\mu\left(\frac{\partial_\nu R}{R}\right)+g_{\mu\nu}\nabla^\alpha\left(\frac{\partial_\alpha R}{R}\right)\right]=\frac{T^{(M)}_{\mu\nu}}{2c_1}\,,
\end{equation}
and do

\begin{equation}
T^{\rm (vac)}_{\mu\nu}\equiv\left(c_0-c_2R\right)g_{\mu\nu}+2c_2\left[\nabla_\mu\left(\frac{\partial_\nu R}{R}\right)-g_{\mu\nu}\nabla^\alpha\left(\frac{\partial_\alpha R}{R}\right)\right]\,,\label{eq:Tvac2}
\end{equation}
\begin{equation}
G_{\mu\nu}=\frac{T^{(M)}_{\mu\nu}+T^{\rm (vac)}_{\mu\nu}}{2\left[c_1+c_2+c_2\ln\left(\frac{R}{R_0}\right)\right]}\,.\label{eq:modiEi}
\end{equation}
Even if the $c_i$'s are constant and matter is covariantly conserved in this scenario we have a dynamical vacuum and a running $G$ with a mild logarithmic dependence, 

\begin{equation}
   \frac{1}{8\pi G(R)}\equiv 2\left[c_1+c_2+c_2\ln\left(\frac{R}{R_0}\right)\right] \,.
\end{equation}
The conservation equation that relates the variation of the two can be obtained straightforwardly from $\nabla^\mu A_{\mu\nu}=0$,

\begin{equation}
\nabla^\mu T_{\mu\nu}^{\rm (vac)}=G^{\mu}_\nu \frac{\partial_\mu G^{-1}}{8\pi}\,.
\end{equation}
The form of the effective gravitational coupling and the vacuum energy-momentum tensor is similar to those in the so-called type-II RRVM \cite{rvmtens,rvmtens2}, in which $T^{(\rm{vac})}_{\mu\nu}=\rho_{\rm vac}(R)g_{\mu\nu}$, with $\rho_{\rm vac}(R)$ a constant plus a linear term in the Ricci scalar $R$. In the case at hand, though, the vacuum EoS parameter is not equal to $-1$ due to the third term of the rhs of \eqref{eq:Tvac2}, namely the one with the coefficient $c_2$. As a result, the vacuum EoS is not proportional to $g_{\mu\nu}$, see Eq.\,\eqref{eq:wvac} below. This scenario leads to the same cosmology as the one obtained in Eq. \eqref{eq:Tvac}, of course, but now the link with the usual RVMs is more explicit since the QFT calculation also points to a vacuum EoS different from $-1$, see \cite{Moreno-Pulido:2022upl}. From now on we focus on this last setup.


\subsection{Background expressions in StRVM}\label{sec:background}

In order to work with dimensionless parameters we define

\begin{equation}\label{c1c2}
c_1+c_2\equiv \frac{d}{16\pi G_N}\qquad , \qquad c_2\equiv \frac{\nu}{16\pi G_N}\,,
\end{equation}
with $G_N$ the Newton constant and
\begin{align}\label{eq:ddeffG}
d= \frac{G_N}{G(z=0)}
\end{align}
the ratio of $G_N$ and the effective cosmological value of the gravitational coupling at present. The modified Friedmann and pressure equations in a flat FLRW universe read, respectively, 

\begin{equation}
3H^2 =\frac{8\pi G_N}{d\left[1+\frac{\nu}{d}\ln\left(\frac{R}{R_0}\right)\right]}\left[\sum_{i=m,r}\rho_i+c_0-\frac{\nu }{16\pi G_N}\left(R+\frac{6H\dot{R}}{R}\right)\right]\,,
\end{equation}
\begin{equation}
-(3H^2+2\dot{H})=\frac{8\pi G_N}{d\left[1+\frac{\nu}{d}\ln\left(\frac{R}{R_0}\right)\right]}\left[\sum_{i=m,r}p_i-c_0+\frac{\nu R}{16\pi G_N}+\frac{\nu}{8\pi G_N}\left(\frac{\ddot{R}}{R}-\frac{\dot{R}^2}{R^2}+\frac{2H\dot{R}}{R}\right)\right]\,,
\end{equation}
with the dots denoting derivatives with respect to the cosmic time. It is not possible to obtain an analytical solution for the Hubble function. However, we can solve the system perturbatively. The running of the effective gravitational coupling and the vacuum is controlled by $\nu$. We expect this parameter to be much smaller than one in order not to depart excessively from General Relativity and respect, among others, the BBN constraints. Thus, we can use to a good approximation, 

\begin{equation}
3H^2 =\frac{8\pi G_N}{d\left[1+\frac{\nu}{d}\ln\left(\frac{R_L}{R_0}\right)\right]}\left[\sum_{i=m,r}\rho_i+c_0-\frac{\nu}{16\pi G_N}\left(R_L+\frac{6H_L\dot{R}_L}{R_L}\right)\right]\,, \label{eq:Fried}
\end{equation}
\begin{equation}
-(3H^2+2\dot{H})=\frac{8\pi G_N}{d\left[1+\frac{\nu}{d}\ln\left(\frac{R_L}{R_0}\right)\right]}\left[\sum_{i=m,r}p_i-c_0+\frac{\nu R_L}{16\pi G_N}+\frac{\nu}{8\pi G_N}\left(\frac{\ddot{R}_L}{R_L}-\frac{\dot{R}_L^2}{R_L^2}+\frac{2H_L\dot{R}_L}{R_L}\right)\right]\,,\label{eq:press}
\end{equation}
where $R_L$ and $H_L$ are the Ricci scalar and the Hubble function at leading order, {\it i.e.} those obtained by solving the system with $\nu=0$. The analytical expressions of the energy densities in terms of the scale factor or the redshift are known in this model. They take the same form as in the $\Lambda$CDM, since matter is covariantly self-conserved, 

\begin{equation}
\dot{\rho}_i+3H(\rho_i+p_i)=0\quad\rightarrow \quad \rho_i=\rho_i^{(0)}a^{-3(1+w_i)}\,,
\end{equation}
with $w_i$ the EoS parameter of the species $i$ and $\rho_i^{(0)}$ its current energy density. For radiation ($i=r$), {\it i.e.}  for photons and massless neutrinos, $w_r=1/3$, whereas for non-relativistic matter ($i=m$), {\it i.e.} baryons and cold dark matter, $w_m=0$. Massive neutrinos have a varying EoS parameter that evolves from $w=1/3$ to $w=0$ when they become non-relativistic.

On the other hand, we know the relation between $R_L$ and the energy-pressure content of the Universe. Hence, we can also express $R_L$ and its time derivatives as a function of $a$ or $z$ very easily, 

\begin{equation}
R_L = \frac{-8\pi G_N}{d}(T^{(M)}+4c_0)=\frac{-8\pi G_N}{d}(4c_0+\rho_m)\,,\label{eq:RL}
\end{equation}
\begin{equation}
\dot{R}_L= \frac{-8\pi G_N}{d}\dot{\rho}_m=\frac{24\pi G_N}{d}H_L\rho_m=\frac{24\pi G_N}{d}\rho_m\left[\frac{8\pi G_N}{3d}(\rho_m+\rho_r+c_0)\right]^{1/2}\,,
\end{equation}
\begin{equation}
\ddot{R}_L=-96\left(\frac{\pi G_N}{d}\right)^2\rho_m\left(2c_0+3\rho_m+\frac{10}{3}\rho_r\right)\,.\label{eq:RLpp}
\end{equation}
%

%

%
Substituting these expressions in \eqref{eq:Fried} and \eqref{eq:press} we find $H(a)$ and its time derivative at first order in $\nu$. Higher-order corrections to these expressions can be computed straightforwardly, but we neglect them here, since their contribution is very small.

In what follows it is useful to define the dimensionless  parameter
\begin{align}\label{epsdef}
\epsilon\equiv \frac{\nu}{d} = \frac{c_2}{c_1+c_2}\,,
\end{align}
where we used \eqref{c1c2}. Both, the numerators and denominators of \eqref{eq:Fried} and \eqref{eq:press} are sensitive to this ratio, which controls the running of the vacuum and $G$. We naturally expect $\epsilon$ to be close to zero and $d$ close to one in order not to depart excessively from the $\Lambda$CDM framework.

Perhaps it is worth to remark that in the supergravity case \eqref{effactionl3}, discussed in section \ref{sec:preinfl}, if one ignores the gravitino condensate ($\sigma_c \to 0$), so as to resemble qualitatively the situation encountered in modern eras, in which gravitinos have decoupled from the spectrum, being very heavy, the coefficients $c_1$ and $c_2$, including one-loop graviton corrections assume the form:
\begin{align}\label{c12}
c_1 - c_2\, {\rm ln}\Big(\frac{H_0^2}{\mu_\tau^2}\Big) &=\frac{1}{2\kappa^2} \Big[1 + \frac{1}{2}\, \kappa^4 \, f^2 \,(0.083 + 0.049 \, {\rm ln}\Big(\frac{\mu_\tau^2}{3\kappa^2\, f^2}\Big))\Big]\,\qquad 
\Rightarrow \nonumber \\
c_1 - c_2\, {\rm ln}\Big(\kappa^2\, H_0^2\Big) &=\frac{1}{2\kappa^2} \Big[1 + \frac{1}{2}\, \kappa^4 \, f^2 \,(0.083 - 0.049 \, {\rm ln}\Big(3\kappa^4\,f^2\Big))\Big] \quad , \quad c_2 = - 0.0045\, \kappa^2 \, f^2 < 0\,,
\end{align}where $-\kappa^2 f^2 <0$ denotes the bare (anti-de-Sitter-type) cosmological constant, as required by 
(local) supersymmetry,  and in the second line of \eqref{c12} we have taken into account ({\it cf.} discussion in section \ref{sec:preinfl}) that the renormalization-group scale $\mu_\tau$ is set~\cite{houston1,houston2} equal to the reduced Planck mass $\kappa^{-1}$ for the dynamical breaking of local supersymmetry, {\it cf.} \eqref{muPl}. The reader should observe that the coefficient $c_2 < 0$ independently of the renormalization group scale $\mu_\tau$.
On the other hand, for perturbative quantum gravity, one may ensure $c_1 > 0$, as required by unitarity, by imposing  
\begin{align}\label{1kappa}1 \gg \kappa^4 f^2 
\end{align} and 
appropriately restricting the argument of the logarithmic term, thus restricting the range of $\mu_\tau$. In our supergravity approach, the choice \eqref{muPl} satisfies this requirement, and thus, in this example, $c_1 > |c_2| > 0$, and therefore the parameter
$\epsilon$ \eqref{epsdef} is negative, independently of the scale $\mu_\tau$. Moreover, as a consequence of \eqref{1kappa}, we have 
\begin{align}\label{orderepssugra} |\epsilon|
~\simeq ~0.009 \,  \kappa^4 f^2 \ll \,1\,.
\end{align}
 In non-supersymmetric quantum gravity models, the bare cosmological constant could be positive, {\it i.e.} one would face a situation in which $f^2 \to -\mathcal E_0 < 0$, and thus $\epsilon >0 $
in such cases. In what follows below, we shall therefore consider a generic phenomenological analysis, including both cases for the parameter $\epsilon$.
 We also stress that upon incorporation of background matter effects,
such as, e.g., cosmic electromagnetic background fields as in \cite{bms1}, 
the coefficient $c_1$ receives additional contributions from such effects. 
Thus $c_1$ and $c_2$ can be considered in practice as independent parameters, and may therefore be fitted as such. This will be used in the subsequent phenomenological analysis.

Using \eqref{epsdef} together with \eqref{eq:RL}-\eqref{eq:RLpp}, the background equations read,

\begin{equation}\label{eq:Fr1}
3H^2= 8\pi G \left(\rho_r+\rho_m+c_0+\frac{\epsilon}{2}(4c_0+\rho_m)+ \frac{3\epsilon(\rho_m+\rho_r+c_0)}{1+\frac{4c_0}{\rho_m}}\right)\,,
\end{equation}

\begin{equation}\label{eq:Fr2}
3H^2+2\dot{H}=-8\pi G\left( p_r-c_0-\frac{\epsilon}{2}(4c_0+\rho_m)+\frac{\epsilon}{2\left(1+\frac{4c_0}{\rho_m}\right)^2}\left[-\rho_m-4c_0+\frac{4c_0}{\rho_m}\left(2c_0+6\rho_r+5\rho_m\right)\right]\right)\,
\end{equation}
with the gravitational coupling 

\begin{equation}\label{eq:effG}
G = \frac{G_N}{d\left[1+\epsilon\ln\left(\frac{4c_0+\rho_m}{4c_0+\rho_m^{(0)}}\right)\right]}\,.
\end{equation}
The vacuum EoS parameter takes the following form\footnote{Notice that this expression depends, in turn, on how we have defined the vacuum EMT in Eq. \eqref{eq:Tvac2}. As already discussed, this definition is not unique. Other choices would leave the physics intact, but would change the form of $w_{\rm vac}$.},

\begin{figure}[t!]
\begin{center}
 \includegraphics[width=6.5in, height=2.5in]{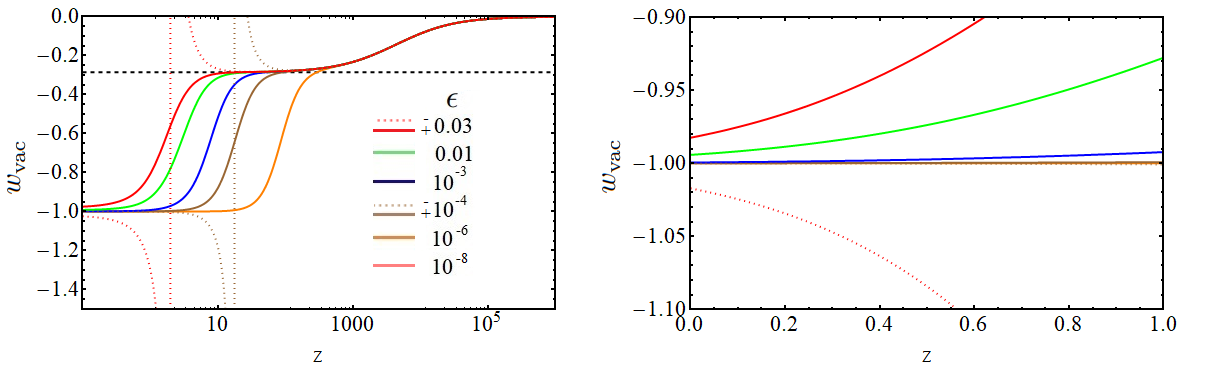}
 \caption{Vacuum EoS parameter $w_{\rm vac}$, Eq. \eqref{eq:wvac}, obtained for different values of $\epsilon$, using $\Omega_m^{(0)}=0.3$ and $\Omega_r^{(0)}=8\cdot 10^{-5}$. The black dashed line in the left plot corresponds to $w_{\rm vac}=-2/7$. In the right plot we zoom in the redshift range $z\in[0,1]$ in order to better grasp the differences between the various curves at low redshifts.}\label{fig:wvac}
\end{center}
\end{figure}

\begin{equation}
w_{\rm vac}=-1+\frac{\frac{3\epsilon\rho_m}{\rho_m+4c_0}(\rho_m+\rho_r+c_0)+\frac{\epsilon\rho_m^2}{2(\rho_m+4c_0)^2}\left[-\rho_m-4c_0+\frac{4c_0}{\rho_m}(5\rho_m+6\rho_r+2c_0)\right]}{c_0+\frac{\epsilon}{2}(\rho_m+4c_0)+\frac{3\epsilon\rho_m}{\rho_m+4c_0}(\rho_m+\rho_r+c_0)}\,.\label{eq:wvac}
\end{equation}
Deep enough in the radiation-dominated epoch (RDE), $w_{\rm vac}\to 0$.  If the condition $|\epsilon\rho_m|\gg c_0$ is fulfilled when radiation becomes negligible compared to non-relativistic matter, {\it i.e.} at $z\sim \mathcal{O}(10^2)$,  $w_{\rm vac}\to -2/7\approx -0.286$ in the matter-dominated era (MDE). Considering typical values of the ratio $c_0/\rho_m^{(0)}\sim 7/3$ we find that this happens if $|\epsilon|\gg 10^{-6}$. Finally,  when $c_0\gg|\epsilon\rho_m|$, $w_{\rm vac}\to -1$ from above (quintessence-like) if $\epsilon>0$ or from below (phantom-like) if $\epsilon<0$. In the negative-$\epsilon$ case, there is  a vertical asymptote at the redshift at which the denominator of Eq. \eqref{eq:wvac} is equal to zero. In Fig. \ref{fig:wvac} we show the shape of $w_{\rm vac}$ for different values of $\epsilon$. The larger the value of $|\epsilon|$, the faster the departure from $w_{\rm vac}=-1$ at low redshifts. For sufficiently small values of $\epsilon>0$, the transition happens directly from $w_{\rm vac}=-1$ to $w_{\rm vac}=0$, with no intermediate plateau at $w_{\rm vac}\sim -2/7$.

It is important to notice that despite the significant deviations of the vacuum EoS parameter from -1 at high redshifts, the EoS of the total cosmological fluid, $w_{\rm tot}=p_{\rm tot}/\rho_{\rm tot}$, remains very close to the $\Lambda$CDM one if $|\epsilon|\ll 1$. In the MDE $w_{\rm vac}=-\epsilon$, and in the RDE $w_{\rm vac}=\frac{1}{3}-\epsilon$.


\subsection{Linear perturbations in StRVM}\label{sec:perturbations}

We compute in Appendix \ref{sec:appendixPert} the perturbed Einstein equations at linear order in the synchronous gauge, that is upon considering the line element $ds^2=a^2[d\tau^2-(\delta_{ij}+h_{ij})dx^idx^j]$. We refer the reader to this appendix for technical details of the computation and the iterative method employed to solve the system of equations. Let us start discussing now the range of values of $\epsilon$ for which the linearly perturbed equations remain reasonably close to the $\Lambda$CDM ones. In order to study this we take for illustrative purposes the perturbed part of the 00 component of Einstein's equations in momentum space, which reads,

\begin{equation}\label{eq:hprime}
\mathcal{H}h^\prime-2\eta k^2 \, =\frac{8\pi G_N a^2\delta\rho}{d\left[1+\epsilon\ln\left(\frac{R_L}{R_0}\right)\right]}+\epsilon\left[\delta R\left(-\frac{a^2}{2}-\frac{k^2}{R_L}+3\mathcal{H}_L\frac{R_L^\prime}{R_L^2}-\frac{3\mathcal{H}_L^2}{R_L}\right)-\frac{h^\prime R_L^\prime}{2R_L}-3\mathcal{H}_L\frac{\delta R^\prime}{R_L}\right]\,.
\end{equation}
The primes denote derivatives with respect to the conformal time $\tau$. The functions $h(\tau,\vec{k})$ and $\eta(\tau,\vec{k})$ are associated to the trace and traceless parts of the perturbed metric in the synchronous gauge, respectively, as defined in \cite{MaBertschinger}. On the other hand, the perturbation of the Ricci scalar takes the form

\begin{equation}
    \delta R= a^{-2}(4\eta k^2-3\mathcal{H}h^\prime-h^{\prime\prime})\,.
\end{equation}
$\epsilon$ is expected to be very small. Thus, almost all the terms inside the brackets of the right-hand side of Eq. \eqref{eq:hprime} are just a small correction to the terms that appear in the left-hand side. Nevertheless, there is one term that can actually compete with the usual terms (those that would survive if $\epsilon$ was set to 0) for sufficiently small scales (large $k$'s). This term is the following,

\begin{equation}\label{eq:conflictTerm}
\epsilon k^2\frac{\delta R}{R_L} = \epsilon k^2 \frac{a^{-2}}{R_L}(4\eta k^2-3\mathcal{H}h^\prime-h^{\prime\prime})\,.
\end{equation} 
We can compare, for instance, the first term in the rhs of \eqref{eq:conflictTerm} with the second term of the lhs of Eq. \eqref{eq:hprime}. It is clear that \eqref{eq:conflictTerm} cannot be treated as a small perturbation unless the condition

\begin{equation}\label{eq:comp}
|\epsilon|\ll \left(\frac{ a \, H}{k}\right)^2
\end{equation}
is fulfilled.  The larger the value of $|\epsilon|$, the smaller the allowed wave numbers $k$ and the larger the scales at which this happens. We want that the above condition is fulfilled, at least, at the typical non-linear scales in $\Lambda$CDM, {\it i.e.} $k_{nl}\gtrsim 0.1$ Mpc$^{-1}$, in order not to spoil the correct description of the CMB and large-scale structure (LSS) observables. This forces us to consider values of $|\epsilon|$ at least $10\%$ smaller than the limiting value $10^{-6}$ (see Fig. \ref{fig:comp}), hence
\begin{align}\label{eps7}
|\epsilon|\lesssim \mathcal{O}(10^{-7})\,.
\end{align} From \eqref{orderepssugra}, then, we observe that for such values one obtains a scale for dynamical supergravity breaking 
\begin{align}\label{f7}
\sqrt{|f|} \gtrsim 10^{-5/4}\, \kappa^{-1}\sim 10^{17}\, {\rm GeV}\,, 
\end{align}
which is sufficiently close to Planck scale to provide support to the pre-RVM-inflationary scenario of \cite{ms1,ms2}. 
We find this estimate rather remarkable. The microscopic dynamics of the stringy RVM involves a pre-inflationary phase characterised by dynamical supergravity breaking at scales much higher than the RVM inflationary scale, which the data place near the GUT scale. Therefore, the range of \eqref{f7}, which ensures, according to our analysis here, that the model alleviates both tensions, is perfectly natural. In fact, it could even be conceived as a prediction of the model, if one reverses the logic. That is, assuming such high supergravity breaking scales, one can arrive at a prediction of the magnitude of the parameter $|\epsilon|$, which enters the data tension phenomenology.

We next remark that when the condition \eqref{eq:comp} is not fulfilled, the perturbation expansion simply breaks down and our linearly perturbed equations become unreliable\footnote{Similar problematic terms can be found in the other perturbed Einstein's equations as well, see Eqs. \eqref{eq:pert1}-\eqref{eq:pert4} in Appendix \ref{sec:appendixPert}.}. Thus, non-linear corrections would be needed in order to solve the system in a proper way at these scales.
\color{black}
This can be seen more explicitly by studying the evolution of the matter perturbations at subhorizon scales ($k^2/\mathcal{H}^2\gg1$). From the matter-dominated epoch onwards, we can approximate Eqs. \eqref{eq:hprime} and \eqref{eq:pert4} as follows,

\begin{figure}[t!]
\begin{center}
 \includegraphics[width=3.5in, height=2.5in]{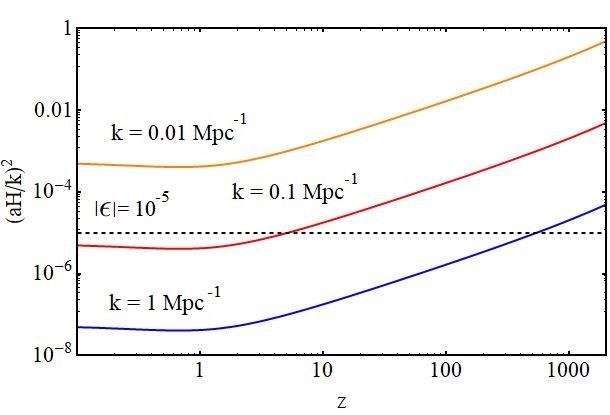}
 \caption{Plot of $(a H/k)^2$ for different comoving wave numbers $k$. To estimate this quantity we have used the Hubble function in the $\Lambda$CDM, with $H_0=70$ km/s/Mpc, $\Omega_m^{(0)}=0.3$ and $\Omega_r^{(0)}=8\cdot 10^{-5}$. This is to assess the range of scales at which the departures from the standard matter growth in the $\Lambda$CDM are kept under control, given a value of $|\epsilon|$ and the condition \eqref{eq:comp}. If we consider a sufficiently large $|\epsilon|$ we violate the condition \eqref{eq:comp} at linear scales and low redshifts. For instance, for $|\epsilon|=10^{-5}$ and $k=0.1$ Mpc$^{-1}$ the condition \eqref{eq:comp} is not fulfilled at $z\lesssim 10$.}\label{fig:comp}
\end{center}
\end{figure}

\begin{equation}
\mathcal{H}h^\prime-2\eta k^2 = \frac{8\pi G_N a^2\delta\rho_m}{d\left[1+\epsilon\ln\left(\frac{R_L}{R_0}\right)\right]}-\epsilon\frac{k^2}{R_L}\delta R\,,
\end{equation}
\begin{equation}
-h^{\prime\prime}-2\mathcal{H}h^\prime+2\eta k^2 =\epsilon\frac{2k^2}{R_L}\delta R\,.
\end{equation}
Here we have neglected the contribution of radiation to the total density and pressure perturbations, since this is a good approximation in the MDE. On the other hand, we have $\delta_{\rm dm}^\prime =-h^\prime/2$ and $\delta_{\rm b}^\prime =-h^\prime/2$, where $\delta_i=\delta\rho_i/\bar{\rho}_i$ is the density contrast of the matter species $i$. Using these results we finally obtain the equation of the baryon and cold dark matter density contrasts,

\begin{equation}\label{eq:denscon}
\delta_i^{\prime\prime}+\mathcal{H}\delta_i^\prime-\frac{4\pi G_N a^2 (\rho_{\rm b}\delta_{\rm b}+\rho_{\rm dm}\delta_{\rm dm})}{d\left[1+\epsilon\ln\left(\frac{R_L}{R_0}\right)\right]}\left(\frac{1+\frac{4\epsilon k^2}{a^2|R_L|}}{1+\frac{3\epsilon k^2}{a^2|R_L|}}\right)=0\,,
\end{equation}
with $i={\rm b},{\rm dm}$. We have done $R_L=-|R_L|$, since $R_L<0$ in our sign convention, see Eq. \eqref{eq:RL}. In this way it will be easier to study later on the r\^ole played by the sign of $\epsilon$. The last equation takes the following form in terms of the scale factor, 

\begin{figure}[t!]
\begin{center}
 \includegraphics[width=7.0in, height=5.0in]{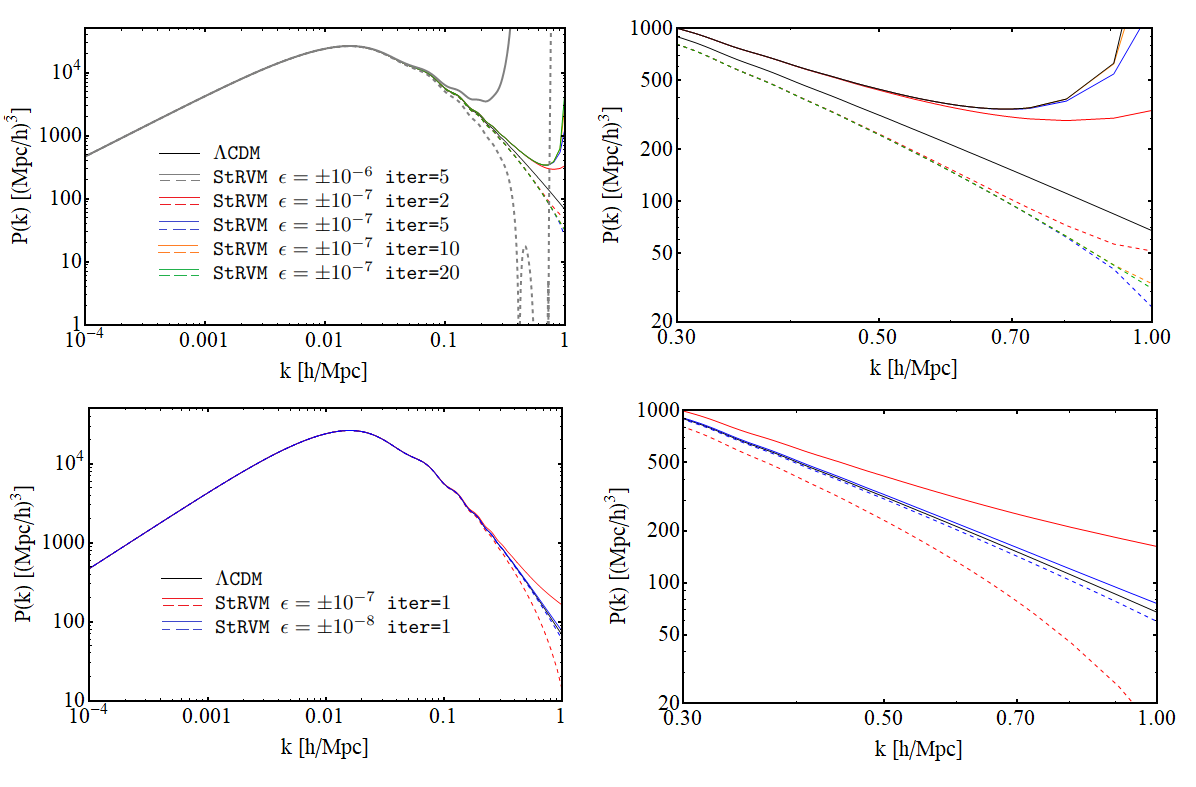}
 \caption{{\it Upper plots:} Matter power spectrum obtained with different values of $\epsilon$ and applying the iterative approach described in Sec. \ref{sec:iterative}, which is implemented in our modified version of \texttt{CLASS}, {\it cf.} Sec. \ref{sec:DataMethod}. We use in all cases the same primordial power spectrum, $d=1$, and the same matter and radiation energy densities. The differences in $\bar{\rho}_{\rm vac}$ are completely negligible for these small $\epsilon$'s. Hence, the values of $h$ are essentially the same, up to a tiny correction of order $\epsilon$. Values $\epsilon\sim \mathcal{O}(10^{-6})$ break the well-tested linearity of $P(k)$ at scales $k\gtrsim 0.1$ Mpc$^{-1}$ and, hence, are unacceptable. The plot on the right is a zoom in the region $k\in[0.3,1]$ $h$Mpc$^{-1}$, which allows us to check that our iterative approach achieves convergence and delivers an output fully in accordance with our analytical study of Sec. \ref{sec:perturbations}. It introduces, though, non-linear effects at the scales at which \eqref{eq:comp} is violated, namely at $k\lesssim 0.2$ $h$Mpc$^{-1}$ if $|\epsilon|=10^{-6}$ and $k\lesssim 0.7$ $h$Mpc$^{-1}$ if $|\epsilon|=10^{-7}$; {\it Lower plots:} Corrected power spectra obtained with $\epsilon=\pm10^{-7},\pm10^{-8}$ by performing only one iteration in order to remove non-linear effects. See the discussion in the main text.}\label{fig:Pk_eps}
\end{center}
\end{figure}

\begin{equation}\label{eq:densSF}
\frac{d^2\delta_i}{da^2}+\left(\frac{3}{a}+\frac{1}{H}\frac{dH}{da}\right)\frac{d\delta_i}{da}-\frac{3\Omega_m}{2a^2}\delta_m \left(\frac{1+\frac{4\epsilon k^2}{a^2|R_L|}}{1+\frac{3\epsilon k^2}{a^2|R_L|}}\right)=0\,,
\end{equation}
with $\Omega_m=\rho_m/\rho_c$ and $\rho_c=\rho_r+\rho_m+\rho_{\rm vac}$ the total energy density. As expected, in the limit $\epsilon\to 0 $ and $d\to 1$ we recover the equation of the $\Lambda$CDM. The effective gravitational coupling that controls the clustering of matter depends on both, the scale factor and $k$, 

\begin{equation}\label{eq:Geff}
G_{\rm eff}(k,a) = \frac{G_N}{d\left[1+\epsilon\ln\left(\frac{R_L}{R_0}\right)\right]}\left(\frac{1+\frac{4\epsilon k^2}{a^2|R_L|}}{1+\frac{3\epsilon k^2}{a^2|R_L|}}\right)=G(a) \left(\frac{1+\frac{4\epsilon k^2}{a^2|R_L|}}{1+\frac{3\epsilon k^2}{a^2|R_L|}}\right) \,,
\end{equation}
where $G(a)$ is given by Eq. \eqref{eq:effG}. In the safe region in which the condition \eqref{eq:comp} is fulfilled, we can Taylor expand Eq. \eqref{eq:Geff} around $\epsilon k^2/|a^2R_L|=0$,

\begin{equation}\label{eq:Geff2}
    G_{\rm eff}(k,a) = \frac{G_N}{d}\left(1+\frac{\epsilon k^2}{a^2|R_L|}\right) +\mathcal{O}\left(\frac{\epsilon^2 k^4}{a^4|R_L|^2}\right) \,.
\end{equation}
Negative (positive) values of $\epsilon$ lead to a suppression (enhancement) of the amount of structure in the Universe, even if it is $|\epsilon|\lesssim\mathcal{O}(10^{-7})$ and, therefore, too small to change significantly the matter energy fraction $\Omega_m$. From Eq. \eqref{eq:densSF} one can also see that, for fixed energy densities, $d$ is fully degenerated with $\epsilon$ in that equation, since the ratio $\epsilon/|R_L|$ is sensitive to the parameter $\nu=\epsilon d$, {\it cf.} formula \eqref{eq:RL}. This means that if $d\lesssim 1$ (for a fixed $ \epsilon> 0$) matter clusters less efficiently at subhorizon scales, whilst if $d\gtrsim 1$ (again, for a fixed $\epsilon> 0$) the aggregation of matter speeds up. This degeneracy can be broken using {\it e.g.} data on $H(z)$. Contrary to $\epsilon$, whose values are very restricted by the LSS data, the parameter $d$ can in principle still deviate from $1$ by $1-10\%$ at cosmological scales without spoiling the description of data from relative cosmic distance indicators like supernovae of Type Ia (SNIa) and baryon acustic oscillations (BAO), respecting also the location of the CMB peaks. This is simply because the ratios of cosmic distances do not depend on $d$\footnote{This also happens in the Brans-Dicke model with a cosmological constant for sufficiently small values of the Brans-Dicke parameter \cite{BDLCDM_CQG}, and also in the type-II RRVM for a slow enough running of the vacuum \cite{rvmtens,rvmtens2}. See Sec. 3 of \cite{BDLCDM_CQG} for further details.}.

The evolution of the matter density contrast at the scales that violate the condition \eqref{eq:comp} is more involved. One can see that if $\epsilon k^2/|a^2R_L|\gg1$, for $\epsilon>0$, the growing mode during the MDE reads $\delta_m\sim a^{\alpha}$, with $\alpha=(1+\sqrt{33})/4\approx 1.7$. Thus, there is a huge enhancement of the matter perturbations compared to the $\Lambda$CDM, where $\delta_m\sim a$. This obviously leads to the non-linear evolution already mentioned before, which basically makes the power spectrum to blow up, as we show in our Fig. \ref{fig:Pk_eps}. If $\epsilon<0$, instead, $G_{\rm eff}$ (Eq. \eqref{eq:Geff}) manifests a weird behavior. It cancels when $\epsilon\to a^2R_L/(4k^2)$ and diverges when $\epsilon\to a^2R_L/(3k^2)$. This fact is imprinted in the matter spectrum, of course, see the upper left plot of Fig. \ref{fig:Pk_eps}. 

 These rare features appear typically at strongly non-linear scales $k\gtrsim 1$ Mpc$^{-1}$ if $|\epsilon|\lesssim\mathcal{O}(10^{-7})$. However, even for these small values of $|\epsilon|$, they can still have a residual (but non-zero) impact on linear quantities like the root-mean-square ({\it rms}) of mass fluctuations at scales of $R_8=8h^{-1}$ Mpc, $\sigma_8$, {\it cf.} the upper right-most plot of Fig. \ref{fig:Pk_eps}. It is important to remove the non-linear effects that are contaminating our theoretical estimation of the linear matter power spectrum at $k\gtrsim 0.5-1$ Mpc$^{-1}$. We do so by considering only one iteration in the iterative method that we apply to solve the system of linear perturbation equations, {\it cf.} Appendix \ref{sec:iterative}. Proceeding in this way we manage to describe very well the power spectrum in the region of $k$'s where \eqref{eq:comp} holds, while keeping under control its amplitude in the region where \eqref{eq:comp} is violated, effectively decoupling the contribution of non-linearities in the calculation.

   For the small values of $|\epsilon|$ under consideration the running of $G$ and the vacuum energy density and pressure are completely negligible, so there is no effect at the background level in StRVM, apart from a pure renormalization of $G$ controlled by the parameter $d$, which can be significant. Nevertheless, as already seen in Fig. \ref{fig:Pk_eps}, even very small values of $\epsilon$ can have a non-negligible impact on the shape of the linear matter power spectrum. We have checked that for a fixed expansion history and primordial power spectrum, an $|\epsilon|\sim \mathcal{O}(10^{-7})$  induces $\mathcal{O}(1\%)$ changes in $\sigma_8$, compared to the $\Lambda$CDM, which is of the same order as the uncertainty obtained for this parameter in CMB studies and can be significant concerning the $\sigma_8$ or $S_8$ tension. Values of $|\epsilon|\sim \mathcal{O}(10^{-8})$, instead, modify $\sigma_8$ only by a $\mathcal{O}(0.1\%)$, much below the sensitivity of current data. As explained in the caption of Fig. \ref{fig:Pk_eps}, larger values of $|\epsilon|$ are unacceptable because they introduce important non-linearities at the very well-tested (mainly linear) BAO scale.  This observation motivates us to use the prior $\epsilon\in[-10^{-7},10^{-7}]$ in our fitting analysis. It is important to notice that in this range of $\epsilon$ the modifications of $P(k)$ happen already at mildly non-linear scales, $k\gtrsim 0.2$ $h$Mpc$^{-1}$, hardly proven by the CMB, {\it cf.} Fig. 19 in \cite{AghanimOverview}. On the other hand, it is not clear how to use galaxy clustering data from redshift-space distortions (RSD) or peculiar velocities to constrain our model, since these measurements are performed by the various galaxy surveys assuming a fiducial $\Lambda$CDM model and, hence, taking the scale-independence of the matter density contrast and the growth factor $f=d\ln\delta_m/d\ln a$ for granted. Thus, the use of these data could induce a bias in our analysis, since in StRVM these quantities depend on $k$ at subhorizon scales. Finally, weak lensing data could also allow us to further constrain $\epsilon$, in principle, but this requires a good theoretical control of the non-linear evolution of matter perturbations in the model, which is out of the scope of this paper\footnote{In the non-linear regime, we expect a potential screening mechanism entering into play and a change (running) of the parameters in the model $(d,\epsilon)$ with respect to the cosmological values. Here we assume that such a screening exists and works fine.}. Therefore, we take a conservative approach and avoid the use of RSD, peculiar velocities and weak lensing data in this study. We will employ the CMB likelihoods from {\it Planck} and cosmological  background data to constrain the StRVM. Due to the aforementioned reasons, we do not expect these datasets to be able to tighten the constraints on $\epsilon$ beyond those set by the prior. 
   
\begin{figure}[t!]
\begin{center}
 \includegraphics[width=7in, height=5.0in]{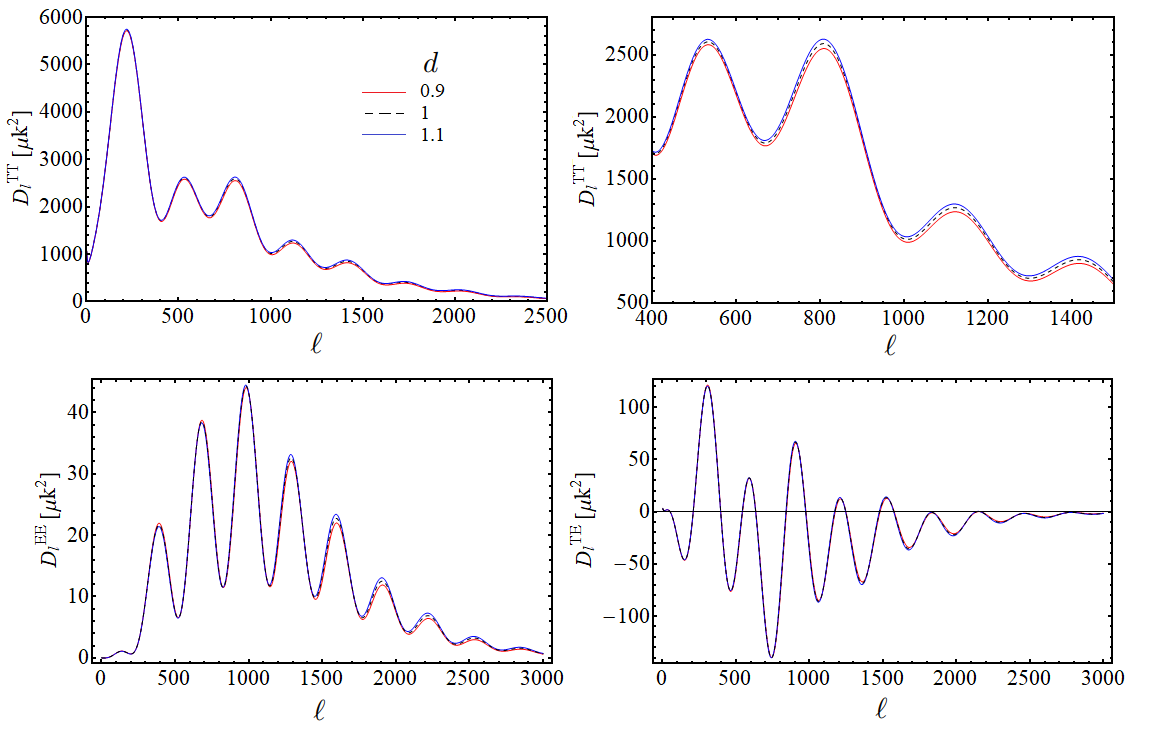}
 \caption{{\it Upper plots:} CMB temperature power spectra obtained with different values of the parameter $d$, using in all cases $\epsilon=0$ and the same optical depth to reionization, same primordial power spectrum and energy densities of the various species. If $d<1$ ($d>1$) there is a suppression (enhancement) of the amplitude, which is more evident at large multipoles, see the right plot. For values of $d$ relatively close to $1$, as those explored in this figure, the angular location of the peaks remain essentially unaffected \cite{BDLCDM_CQG}. {\it Lower plots:} CMB EE polarization spectrum (on the left) and the ET cross-correlation spectrum (on the right), obtained for the same setup as in the upper plots. See the comments in the main text.}\label{fig:Cld}
\end{center}
\end{figure}
   
   As already discussed, $\epsilon$ does not alter the background evolution nor the CMB spectra for the range of values covered by our prior. Hence, in order to discuss now the r\^ole played by the parameter $d$, we can just set $\epsilon=0$ for convenience. In Fig. \ref{fig:Cld} we show the temperature CMB spectrum obtained with $d=1,\,0.9$ and $1.1$, by using the same energy densities of the various species, reionization depth and primordial power spectrum. This translates into a constant cosmological  $G$ equal to the standard value $G_N$, $1.1G_N$ and $0.9G_N$, respectively. Values of $d$ below (above) one lead to a decrease (increase) of the amplitude of the spectra, more conspicuously at large multipoles. The larger is the effective Newton constant the faster the expansion and, hence, the longer the recombination time. This increases the photon diffusion at small angular scales \cite{Silk1968}, which explains the suppression of the $D_l^{\rm TT}$'s for $d<1$ \cite{Zahn2002}. See also \cite{Galli2009,Salvado2015,BDLCDM_CQG}. We will constrain $d$ using CMB data and measurements on cosmic chronometers and $H_0$. The latter provide information about the absolute distance scale of the Universe, instead of relative distances like  uncalibrated SNIa or BAO, which are useful to constrain the parameter space of the theory, of course, but do not constrain $d$ when used alone. The changes induced in the CMB TT spectrum by $d$ can be compensated at large extent by shifts of the spectral index $n_s$ \cite{Zahn2002,Galli2009,Salvado2015,BDLCDM_CQG}. CMB data introduce a strong anti-correlation between these two parameters, which is partially broken by CMB polarization data, see \cite{Zahn2002} and Fig. \ref{fig:Cld}. This anti-correlation is welcome because might allow $d$ to depart from 1 in a non-negligible way. Values of $d<1$ may alleviate the Hubble tension, since they enhance $H(z)$ by a factor that does not change throughout the cosmic expansion. It is also interesting to note that for $\epsilon=0$ and at subhorizon scales the growth in the MDE and late-time Universe is not affected by $d$, as it can be easily seen from Eq. \eqref{eq:densSF}. Notice, however, that $P(k)$ is clearly sensitive to this parameter, since the moment at which the modes become subhorizon and other many features like the location of the peak depend on $d$, {\it cf.} Fig. \ref{fig:Pk_d}.  As explained in \cite{BDLCDM_CQG}, the value of $\sigma_8$ remains unmodified in this case under changes of $d$, despite the changes induced in $P(k)$. This is because the latter are exactly compensated by shifts in the scale $R_8=8/h$ Mpc$^{-1}$ at which the window function peaks\footnote{Remember that $\sigma_8$ does not characterize the amount of clustering at a fixed scale if $h$ is not kept constant \cite{Sanchez2020}. This is also another argument against the use of RSD data in this work.}. Hence, in principle, RSD data, which constraints the combination $f\sigma_8$, cannot constrain $d$ neither. Notice, though, that measurements of the {\it rms} of mass fluctuations at a fixed scale $R$ (independent from $h$), or the shape of the matter power spectrum could be employed to constrain the StRVM. 

   The StRVM has a very interesting phenomenology and may have something to say concerning the cosmological tensions. A value of $d\sim 0.90-0.95$ could let us alleviate the existing Hubble tension between the distance ladder measurement by SH0ES \cite{RiessH02022} and the determination from {\it Planck}'s CMB data \cite{planck}, and also produce a smaller amplitude of the power spectrum (see again Fig. \ref{fig:Pk_d}). In addition, a negative $\epsilon$ of order $10^{-7}$ may also contribute to suppress the structure formation processes in the Universe (cf. Fig. \ref{fig:Pk_eps}).

\begin{figure}[t!]
\begin{center}
 \includegraphics[width=7.0in, height=2.5in]{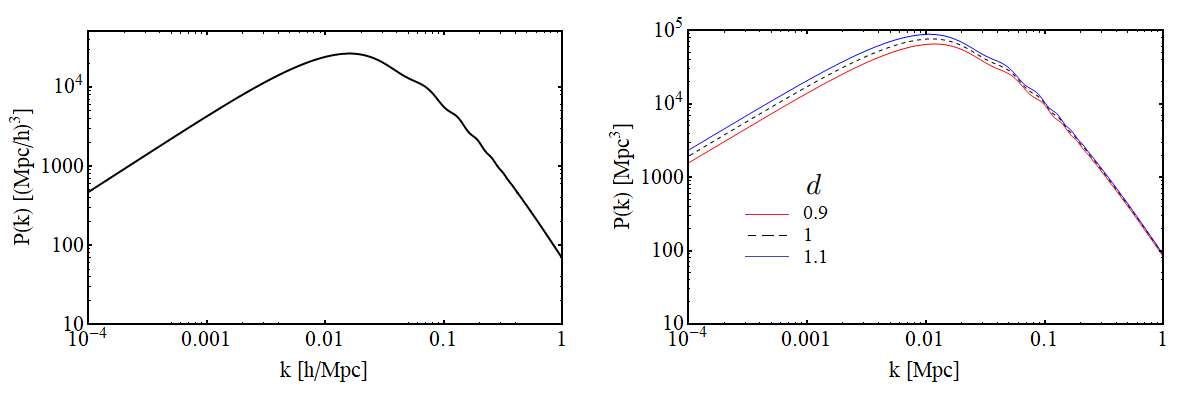}
 \caption{{\it Left plot}: Linear matter power spectrum in units of (Mpc/$h)^{3}$ obtained for the $\Lambda$CDM and two StRVM models with  $d=0.9$ and $d=1.1$. We use the same primordial power spectrum and energy densities, setting in both cases $\epsilon=0$. There is no difference between the three curves, but this is just due to the units employed in this plot; {\it Right plot:} The same, but now in units of Mpc$^3$. The differences between the three models become now evident. The amplitude of $P(k)$ is lower for values of $d<1$. The power spectrum at present reads, $P(k)=(k/\mathcal{H}_0)^4T^2(k)\delta^2_H(k)$, with $T(k)$ the transfer function and $\delta^2_H$ an amplitude which is directly related to the primordial power spectrum \cite{LiddleLyth}. At superhorizon scales $T(k)\approx 1$, so we have $P(k,d=1.1)/P(k,d=0.9)=(1.1/0.9)^2\approx 1.49$, since we have fixed the primordial power spectrum. There are also differences at larger $k$'s, of course. For instance, the location of the peak $k_{\rm eq}$ and the comoving wave number at the horizon crossing $k_{\rm hor}$ scale both as $k_{\rm eq},k_{\rm hor}\propto d^{-1/2}$. }\label{fig:Pk_d}
\end{center}
\end{figure}

\section{Data and methodology}\label{sec:DataMethod}

We have implemented the StRVM model in our own modified version of the Einstein-Boltzmann code \texttt{CLASS} \cite{CLASS1,CLASS2}, which allows us to solve the system of background and linear perturbations equations and compute all the theoretical quantities that are required to confront the model with observations. See Sec. \ref{sec:background} and Appendix \ref{sec:appendixPert} for the relevant expressions in StRVM.

In our fitting analyses we have employed the following datasets: 

\begin{enumerate}
    \item The high- and low-$\ell$ data of the
    CMB temperature and polarization spectra and their cross-correlations measured by the {\it Planck} satellite \cite{planck}. This is the Planck 2018 TT,TE,EE+lowE likelihood, or Planck18 in short. In some of our analyses we exclude the use of the high-$\ell$ polarization data. Some authors have found a moderate inconsistency between this dataset and the CMB temperature anisotropies measured by {\it Planck} \cite{Lin2017,GarciaQuintero2019}, so it is good to study what is the impact of removing the polarization information. We denote the resulting CMB dataset as Planck18(np). Finally, we also incorporate in some cases the CMB lensing by considering the Planck 2018 TT,TE,EE+lowE+lensing likelihood, what we call Planck18(lens).
    
    \item The compilation Pantheon+ of SNIa \cite{Scolnic2022}. It includes 1701 light
curves of 1550 SNIa, ranging in redshift from $z = 0.001$ to $2.26$. In this work we actually use two different SNIa samples. In our main analyses we use the sample with
1624 data points that is obtained upon the removal of the SNIa contained in the host galaxies of SH0ES \cite{RiessH02022,Brout2022}. When we use the full Pantheon+ compilation together with the distances to the host galaxies measured by SH0ES we denote the resulting SNIa dataset as SnIa\_$H_0$. The statistical and systematic uncertainties are included through the corresponding covariance matrix.

    \item The BAO measurements from the galaxy surveys 6dFGS+SDSS MGS \cite{Carter2018}, WiggleZ \cite{Kazin2014}, BOSS DR12 \cite{GilMarin2016}, DES Y3 \cite{Abbott2022} and eBOSS DR16 \cite{Neveux2020,Hou2020,duMasdesBourboux2020}. We have considered the corresponding covariance matrices, accounting for the internal correlations in WiggleZ, BOSS DR12 and eBOSS DR16.
    
    \item 32 data points on $H(z_i)$ from cosmic chronometers in the redshift range $0.07<z<1.965$ obtained with the differential age technique and passively evolving galaxies \cite{JimenezLoeb}. They are provided in references \cite{Jimenez2003,Simon2005,Stern2010,Moresco2012,Zhang2014,Moresco2015,Moresco2016,Ratsimbazafy2017,Borghi2022}. We have taken into account the various sources of statistical and  systematic errors and the existing correlations, as explained in \cite{Moresco2020}.  
    
\end{enumerate}
We have constrained the parameter space of the model making use of the Metropolis-Hastings algorithm \cite{Metropolis,Hastings} that is integrated in the Monte Carlo sampler \texttt{MontePython} \cite{MP1,MP2}. We stop the Monte Carlo routine when the Gelman-Rubin convergence statistic $R-1<5\cdot 10^{-3}$ \cite{R1,R2}. Our Markov chains are analyzed with the Python code \texttt{GetDist} \cite{GetDist}, which is also employed to obtain the marginalized constraints presented in Tables \ref{tab:tableI}-\ref{tab:tableII} and the contour plots of Fig \ref{fig:contours}. In our Monte Carlo routine we perform jumps in the parameter space composed by the parameters $(\omega_b,\omega_{\rm cdm},H_0,\tau_{\rm reio},A_s, n_s)$ and the StRVM parameters $(\epsilon,d)$, using the prior $\epsilon\in[-10^{-7},+10^{-7}]$, as discussed in Sec. \ref{sec:perturbations}. We assume the standard form of the primordial power spectrum in our main analyses, $P_{\mathcal{R}}=A_s(k/k_*)^{n_s-1}$, with the pivot scale $k_* =0.05$ Mpc$^{-1}$.  We obtain as derived parameters the comoving sound horizon at the baryon-drag epoch, $r_d$, $\sigma_8\equiv \sigma(R=8h^{-1}{\rm Mpc})$, $\sigma_{12}\equiv \sigma(R=12{\rm Mpc})$, and the combined quantities \cite{Sanchez2020}

\begin{equation}\label{eq:s8s12}
    S_8=\sigma_8\left(\frac{\Omega_m}{0.3}\right)^{0.5}\qquad {\rm and}\qquad S_{12} = \sigma_{12}\left(\frac{\omega_m}{0.14}\right)^{0.4}\,.
\end{equation}
Let us note that the quantity $\sigma_{12}$ (and the associated $S_{12}$) has been claimed to be  more suitable for the analysis of the growth tension than the more traditional ones $\sigma_{8}$ ($S_{8}$) -- see \cite{{Sanchez2020}} for details-- and for this reason we shall report on both type of parameters \eqref{eq:s8s12} in our analysis.
When the SNIa of the Pantheon+ compilation are used, we also report in our tables the values of their standardized absolute magnitude, $M$.

In addition, in some of our analyses we also explore the effect of the running of the spectral index, $\alpha_s$, and the running of the running, $\beta_s$\footnote{The primordial scalar perturbation spectrum is assumed to take the following power-law form,  $$\mathcal{P}_\mathcal{R}(k)=A_s\left(\frac{k}{k_*}\right)^{n(k)}\,,$$ with $k_*=0.05$ Mpc$^{-1}$ the pivot scale, $$n(k)=n_s-1+\frac{\alpha_s}{2}\ln\left(\frac{k}{k_*}\right)+\frac{\beta_s}{6}\ln^2\left(\frac{k}{k_*}\right)\,,$$ and $$\alpha_s\equiv\frac{dn_s}{d\ln k}\qquad {\rm and}\qquad \beta_s\equiv\frac{d^2n_s}{d\ln k^2}$$ the running of $n_s$ and the running of the running of $n_s$, respectively \cite{planck}.}, to see whether this can help the model to improve the overall description of the data. The EE polarization data break to some extent the strong anti-correlation between $n_s$ and $d$. It is clear from Fig. \ref{fig:Cld} that while $n_s$ can compensate the effects of a change in $d$ in the TT spectrum, it is unable to compensate the changes induced in the EE spectrum at mulipoles $\ell\lesssim 1000$. Here we study whether a running in the spectral index can solve this problem.

\begin{table*}[t!]
\centering
\begin{tabular}{|c ||c | c || c| c|| c| c|    }
 \multicolumn{1}{c}{} & \multicolumn{1}{c}{} & \multicolumn{1}{c}{} & \multicolumn{1}{c}{} & \multicolumn{1}{c}{} \\
\multicolumn{1}{c}{} &  \multicolumn{2}{c}{Planck18} &  \multicolumn{2}{c}{Planck18+SNIa+BAO+CCH}&  \multicolumn{2}{c}{Planck18+SNIa\_$H_0$+BAO+CCH}
\\\hline
{\small Parameter} & {\small $\Lambda$CDM}  & {\small StRVM}& {\small$\Lambda$CDM} & {\small StRVM}& {\small$\Lambda$CDM} & {\small StRVM}
\\\hline
$10^2\omega_b$ &  $2.232\pm 0.015$ & $2.228^{+0.020}_{-0.021}$ & $2.243\pm 0.013$ & $2.249^{+0.017}_{-0.016}$ & $2.257^{+0.012}_{-0.013}$ & $2.279^{+0.013}_{-0.014}$ \\\hline
$\omega_{\rm cdm}$ & $0.1197\pm 0.0014$ & $0.1197\pm 0.0014$  & $0.1181\pm 0.0008$ & $0.1182\pm 0.0008$ & $0.1169^{+0.0007}_{-0.0008}$ & $0.1180\pm 0.0008$ \\\hline
$n_s$ & $0.9627^{+0.0045}_{-0.0043}$ & $0.9613^{+0.0074}_{-0.0072}$ & $0.9664\pm 0.0036$ & $0.9687^{+0.0059}_{-0.0057}$ & $0.9693^{+0.0035}_{-0.0036}$ & $0.9813^{+0.0046}_{-0.0045}$ \\\hline
$\tau_{\rm reio}$ & $0.054^{+0.007}_{-0.008}$ & $0.054\pm 0.008$ & $0.056^{+0.007}_{-0.008}$ & $0.056\pm 0.008$  & $0.058\pm 0.008$ & $0.059^{+0.008}_{-0.009}$\\\hline
$\sigma_{12}$ & $0.805^{+0.010}_{-0.011}$ & $0.809\pm 0.019$ & $0.795\pm 0.008$& $0.794\pm 0.016$ & $0.788\pm 0.008$ & $0.779^{-0.015}_{-0.014}$ \\\hline
$H_0$ & $67.11^{+0.59}_{-0.62}$ & $66.79\pm 1.46$ & $67.80\pm 0.36$ & $68.35^{+1.15}_{-1.11}$ & $68.40\pm 0.33$ & $71.27^{+0.76}_{-0.73}$ \\\hline
$d$ & $-$ & $1.008^{+0.029}_{-0.032}$ & $-$ & $0.986^{+0.024}_{-0.028}$ & $-$ & $0.924\pm 0.017$\\\hline\hline
$r_d$ & $147.40^{+0.31}_{-0.30}$ & $148.00^{+2.31}_{-2.46}$ & $147.69^{+0.21}_{-0.22}$ & $146.54^{+2.07}_{-2.22}$ & $147.85\pm 0.22$ & $141.52^{+1.42}_{-1.49}$ \\\hline
$M$ & $-$ & $-$ & $-19.430\pm 0.011$ & $-19.413^{+0.037}_{-0.034}$ & $-19.411^{+0.009}_{-0.010}$ & $-19.323^{+0.023}_{-0.022}$ \\\hline
$S_{8}$ & $0.829^{+0.016}_{-0.017}$ & $0.831^{+0.022}_{-0.020}$ & $0.812\pm 0.011$& $0.815^{+0.017}_{-0.018}$ & $0.799^{+0.011}_{-0.010}$ & $0.816^{+0.019}_{-0.017}$\\\hline
$\sigma_8$ & $0.809^{+0.007}_{-0.008}$ & $0.810^{+0.017}_{-0.015}$ & $0.805^{+0.006}_{-0.007}$ &  $0.808^{+0.016}_{-0.015}$ & $0.801^{+0.007}_{-0.008}$ & $0.816^{+0.017}_{-0.015}$
 \\\hline
$S_{12}$ & $0.810^{+0.013}_{-0.014}$ & $0.814^{-0.021}_{-0.020}$ & $0.797\pm 0.009$ & $0.796\pm 0.016$ & $0.786\pm 0.009$ & $0.780^{+0.015}_{-0.014}$  \\\hline
$\chi^2_{\rm min}$ & 2767.97 &   2767.28 & 4255.34 & 4255.30 & 4293.24 & 4273.16
\\\hline
$\Delta{\rm DIC}$ & $-$  & -2.78  & $-$ & -1.68 & $-$ & +17.65
\\\hline
\end{tabular}
\caption{Mean and uncertainties at $68\%$ C.L. of the individual parameters of the $\Lambda$CDM and StRVM models for all the data combinations involving Planck18. As explained in the main text, the constraints on $\epsilon$ are dominated by the prior $\epsilon\in[-10^{-7},+10^{-7}]$. $H_0$ and $r_d$ are expressed in km/s/Mpc and Mpc, respectively. To assess the fitting performance of the models in the last two rows we report  for each analysis the values of $\chi^2_{\rm min}$ and $\Delta {\rm DIC}\equiv{\rm DIC}_{\Lambda{\rm CDM}}-{\rm DIC}_{\rm StRVM}$ \eqref{eq:DIC}. }
\label{tab:tableI}
\end{table*}

Finally, we also study the status of the CMB lensing anomaly in the context of the StRVM. This anomaly is an excess of CMB lensing power in the temperature spectra measured by WMAP \cite{Calabrese2008} and {\it Planck} \cite{planck} under the assumption of the $\Lambda$CDM. To perform this study we rescale the theoretical prediction of the lensing spectrum by the so-called $A_L$ consistency parameter \cite{Calabrese2008}. Significant deviations from $A_L=1$ indicate a tension. In the standard model, this anomaly reaches the $2.5\sigma$ and $2.8\sigma$ c.l. with Planck18(np) and Planck18, respectively \cite{planck}. 

Due to the reasons explained in the preceding section, we believe it is safer to avoid the use of data on $f\sigma_8$ and from weak lensing. We perform our analyses first with Planck18 alone to study the constraining power of the CMB. Then, we combine CMB with background data without including the information from SH0ES, {\it i.e.} Planck18+SNIa+BAO+CCH, and also including it, {\it i.e.} Planck18+SNIa\_$H_0$+BAO+CCH. This will allow us to assess the impact of the cosmic distance ladder measurements on our results. Then, we repeat all the fitting analyses excluding the high-$\ell$ CMB polarization data and, finally, keeping the polarization data and including the CMB lensing information.  

For the sake of comparison, in our tables and figures we present not only the fitting results obtained within the StRVM, but also those obtained within the $\Lambda$CDM framework. Apart from the mean values and uncertainties of the various parameters, we also list in our tables the minimum values of $\chi^2$, {\it i.e.} $\chi^2_{\rm min}$, and the difference between the deviance  information criteria (DIC) \cite{DIC} in the two models, $\Delta{\rm DIC}\equiv{\rm DIC}_{\Lambda{\rm CDM}}-{\rm DIC}_{\rm StRVM}$. DIC penalizes the use of additional parameters. It is defined as

\begin{equation}\label{eq:DIC}
   {\rm DIC} =  \chi^2(\bar{\theta})+2p_D\,,
\end{equation}
with $p_D=\overline{\chi^2}-\chi^2(\bar{\theta})$ the effective number of parameters in the model and $\bar{\theta}$ the mean of the parameters that are left free in the Monte Carlo analysis. A positive difference of DIC implies that the StRVM performs better than the $\Lambda$CDM, whereas negative differences mean just the opposite. If $0 \leq \Delta\textrm{DIC}<2$ we find a weak evidence in favor of StRVM. If  $2 \leq \Delta\textrm{DIC} < 6$ we speak, instead, of positive evidence. If $6 \leq \Delta\textrm{DIC} < 10$ there is strong evidence in favor of StRVM, whilst if  $\Delta\textrm{DIC}>10$ we can conclude that there is very strong evidence supporting our StRVM model.

\begin{table*}[t!]
\centering
\begin{tabular}{|c ||c | c || c| c|| c| c|    }
 \multicolumn{1}{c}{} & \multicolumn{1}{c}{} & \multicolumn{1}{c}{} & \multicolumn{1}{c}{} & \multicolumn{1}{c}{} \\
\multicolumn{1}{c}{} &  \multicolumn{2}{c}{Planck18(np)} &  \multicolumn{2}{c}{Planck18(np)+SNIa+BAO+CCH}&  \multicolumn{2}{c}{Planck18(np)+SNIa\_$H_0$+BAO+CCH}
\\\hline
{\small Parameter} & {\small $\Lambda$CDM}  & {\small StRVM}& {\small$\Lambda$CDM} & {\small StRVM}& {\small$\Lambda$CDM} & {\small StRVM}
\\\hline
$10^2\omega_b$ &  $2.208\pm 0.022$ & $2.211\pm 0.029$  &   $2.222^{+0.020}_{0.019}$  & $2.233\pm 0.022$  & $2.244\pm 0.019$ & $2.261\pm 0.019$ \\\hline
$\omega_{\rm cdm}$ & $0.1201^{+0.0021}_{-0.0022}$ & $0.1199^{+0.0022}_{-0.0021}$ & $0.1178\pm 0.0009$ & $0.1179\pm 0.0009$ & $0.1163^{+0.0008}_{-0.0009}$ & $0.1180\pm0.0009$  \\\hline
$n_s$ & $0.961\pm 0.006$ & $0.962\pm 0.010$ & $0.966\pm 0.004$  & $0.971\pm 0.007$ & $0.970\pm 0.004$ & $0.983\pm 0.005$ \\\hline
$\tau_{\rm reio}$ & $0.052\pm 0.008$ & $0.052\pm 0.008$ & $0.054\pm 0.008$   &   $0.054\pm 0.008$ & $0.056^{+0.007}_{-0.008}$ & $0.056\pm 0.008$ \\\hline
$\sigma_{12}$ & $0.808^{+0.015}_{-0.016}$ & $0.807^{+0.023}_{-0.024}$ &  $0.793\pm 0.009$  &  $0.788^{+0.016}_{-0.017}$    & $0.783^{+0.009}_{-0.008}$ & $0.772^{+0.015}_{-0.014}$  \\\hline
$H_0$ & $66.71^{+0.94}_{-0.92}$ & $67.07^{+2.19}_{-2.26}$ & $67.75^{+0.39}_{-0.41}$   & $69.07^{+1.46}_{-1.43}$  & $68.53^{+0.38}_{-0.37}$ & $72.07^{+0.80}_{-0.87}$  \\\hline
$d$ & $-$ &  $0.995^{+0.041}_{-0.047}$ & $-$ & $0.968^{+0.032}_{-0.036}$ &  $-$ & $0.902\pm 0.019$\\\hline\hline
$r_d$ & $147.57\pm 0.49$ & $147.2^{+3.2}_{-3.6}$ & $148.02^{+0.30}_{-0.31}$ & $145.4^{+2.7}_{-2.9}$  & $148.16^{+0.30}_{-0.31}$ & $140.0\pm 1.6$  \\\hline
$M$ & $-$ & $-$ & $-19.432\pm 0.012$ & $-19.391^{+0.046}_{-0.044}$  & $-19.408\pm 0.011$ & $-19.299^{+0.024}_{-0.025}$ \\\hline
$S_{8}$ & $0.834^{+0.024}_{-0.025}$ & $0.833\pm 0.027$ & $0.808\pm 0.012$ & $0.813\pm 0.018$ & $0.792\pm 0.011$ & $0.816^{+0.018}_{-0.019}$ \\\hline
$\sigma_8$ & $0.808\pm 0.009$ & $0.810^{+0.017}_{-0.016}$ & $0.802\pm 0.012$  &  $0.807^{+0.017}_{-0.016}$  & $0.798^{+0.007}_{-0.008}$ &  $0.815^{+0.017}_{-0.016}$
 \\\hline
$S_{12}$ & $0.813^{+0.019}_{-0.020}$ & $0.811^{-0.026}_{-0.027}$ & $0.793\pm 0.010$   & $0.788^{+0.017}_{-0.018}$  & $0.781\pm 0.010$ & $0.773\pm 0.015$ \\\hline
$\chi^2_{\rm min}$ &  1179.54 & 1178.80  & 2667.78 & 2667.34  & 2704.42 & 2677.84 
\\\hline
$\Delta{\rm DIC}$ & $-$  & -2.12  & $-$ & +0.44 & $-$ & +23.81 
\\\hline
\end{tabular}
\caption{Same as in Table \ref{tab:tableI}, but without including the high-$\ell$ CMB polarization data.}
\label{tab:tableIII}
\end{table*}


\section{Results} \label{sec:results}

We discuss now our fitting results for the $\Lambda$CDM and the StRVM. We display them in Tables \ref{tab:tableI}-\ref{tab:tableII}. They have been obtained making use of the datasets and applying the methodology explained in the preceding section. In Appendix \ref{sec:extraTables} we present some additional tables with the breakdown of the contributions to $\chi^2_{\rm min}$ from each observable, which will prove useful to understand some aspects of the analyses.

The StRVM is characterized by two additional parameters with respect to the vanilla $\Lambda$CDM model, $d$ and $\epsilon$. They control the current value of the gravitational coupling at cosmological scales $G$, and the running of the vacuum and $G$, respectively. Unfortunately, none of the datasets considered in this work is able to improve significantly our prior constraints on $\epsilon$ ($|\epsilon|<10^{-7}$) for the reasons already discussed in Sec. \ref{sec:perturbations}. This is why we do not include this parameter in our tables. The situation regarding $d$, though, is different. When CMB data is used alone in the fitting analysis in any of its variants, {\it i.e.} Planck18, Planck18(np) or Planck18(lens), we find values of $d$ fully compatible with 1 and, hence, also a full compatibility between the measured $G$ and Newton's constant. For instance, we find $d=0.995^{+0.041}_{-0.047}$ with Planck18(np) and $d=1.008^{+0.029}_{-0.032}$ with Planck18. The inclusion of the CMB polarization data allows us to reduce the uncertainties significantly, by  $\gtrsim 30\%$. When, on top of the temperature and polarization data, we also consider the CMB lensing, we get $d=1.010^{+0.029}_{-0.032}$, showing that the CMB lensing does not improve the overall constraining power, since the uncertainties remain essentially the same. The fit values of the standard cosmological parameters remain also very close to those found in the $\Lambda$CDM. Actually, the StRVM only allows for a slight decrease of $\chi^2_{\rm min}$ when only CMB data are employed in the analyses. This explains the small differences of DIC, which lie in all cases in the negative range $-2.8\lesssim\Delta{\rm DIC}\lesssim -1.5$ and, hence, do not point to any statistical preference for the StRVM once we penalize the use of additional parameters in the model. Despite the central values of all the parameters being close to the $\Lambda$CDM ones, there is a non-negligible broadening of their posteriors. In particular, the one of $H_0$, which allows to reach the region $\gtrsim70$ km/s/Mpc at $\sim 2\sigma$ C.L.

\begin{table*}[t!]
\centering
\begin{tabular}{|c ||c | c || c| c|| c| c|    }
 \multicolumn{1}{c}{} & \multicolumn{1}{c}{} & \multicolumn{1}{c}{} & \multicolumn{1}{c}{} & \multicolumn{1}{c}{} \\
\multicolumn{1}{c}{} &  \multicolumn{2}{c}{Planck18(lens)} &  \multicolumn{2}{c}{Planck18(lens)+SNIa+BAO+CCH}&  \multicolumn{2}{c}{Planck18(lens)+SNIa\_$H_0$+BAO+CCH}
\\\hline
{\small Parameter} & {\small $\Lambda$CDM}  & {\small StRVM}& {\small$\Lambda$CDM} & {\small StRVM}& {\small$\Lambda$CDM} & {\small StRVM}
\\\hline
$10^2\omega_b$ & $2.231^{+0.016}_{-0.014}$  & $2.228\pm 0.020$ &  $2.243^{+0.014}_{-0.013}$   &  $2.248^{+0.016}_{-0.017}$  & $2.258\pm 0.013$ & $2.279\pm 0.014$ \\\hline
$\omega_{\rm cdm}$ & $0.1198^{+0.0012}_{-0.0013}$ & $0.1198\pm 0.0012$ & $0.1183\pm 0.0008$ & $0.1184\pm 0.0008$  & $0.1171^{+0.0008}_{-0.0007}$ & $0.1182\pm0.0008$  \\\hline
$n_s$ & $0.962\pm 0.004$ & $0.961\pm 0.007$ & $0.966\pm 0.004$ &  $0.968\pm 0.006$ & $0.969\pm 0.004$ & $0.981^{+0.005}_{-0.004}$ \\\hline
$\tau_{\rm reio}$ & $0.054^{+0.007}_{-0.008}$ & $0.054^{+0.007}_{-0.008}$ &  $0.059^{+0.007}_{-0.008}$  &  $0.059^{+0.007}_{-0.008}$  & $0.062^{+0.007}_{-0.008}$ & $0.062\pm 0.008$ \\\hline
$\sigma_{12}$ & $0.807\pm 0.008$ & $0.810\pm 0.017$ & $0.799\pm 0.007$ & $0.799\pm 0.012$ & $0.793\pm 0.007$ & $0.784^{+0.015}_{-0.012}$  \\\hline
$H_0$ & $67.04^{+0.54}_{-0.55}$ & $66.67^{+1.43}_{-1.42}$ & $67.72\pm 0.35$  & $68.27^{+1.12}_{-1.14}$  & $68.33^{+0.32}_{-0.35}$ & $71.18^{+0.73}_{-0.79}$  \\\hline
$d$ & $-$ & $1.010^{+0.029}_{-0.032}$  & $-$ & $0.987^{+0.025}_{-0.028}$ &  $-$ & $0.924^{+0.016}_{-0.017}$\\\hline\hline
$r_d$ & $147.37^{+0.29}_{-0.27}$ & $148.1^{+2.4}_{-2.5}$ & $147.64\pm 0.21$ & $146.6^{+2.1}_{-2.3}$ & $147.79^{+0.22}_{-0.21}$ & $141.5\pm 1.4$  \\\hline
$M$ & $-$ & $-$ & $-19.432^{+0.010}_{-0.011}$ & $-19.416^{+0.036}_{-0.035}$ & $-19.413\pm 0.010$ & $-19.325^{+0.022}_{-0.023}$ \\\hline
$S_{8}$ & $0.831\pm 0.013$ & $0.833^{+0.019}_{-0.018}$ & $0.817\pm 0.009$ & $0.821\pm 0.016$ & $0.806\pm 0.009$ & $0.822^{+0.018}_{-0.015}$ \\\hline
$\sigma_8$ & $0.810\pm 0.006$ & $0.810^{+0.016}_{-0.014}$ &  $0.808\pm 0.006$ &  $0.812\pm 0.015$  & $0.807\pm0.006$ &  $0.821^{+0.018}_{-0.013}$
 \\\hline
$S_{12}$ & $0.811\pm 0.010$ & $0.815\pm 0.018$ & $0.801\pm 0.008$ & $0.801^{+0.015}_{-0.016}$ & $0.792^{+0.008}_{-0.009}$ & $0.786^{+0.015}_{-0.012}$ \\\hline
$\chi^2_{\rm min}$ & 2775.28  & 2775.22  & 4266.26 & 4307.30 & 4303.35 & 4283.43
\\\hline
$\Delta{\rm DIC}$ & $-$  & -1.52 & $-$ & -2.58 & $-$ & +19.11 
\\\hline
\end{tabular}
\caption{Same as in Table \ref{tab:tableI}, but including the CMB lensing data.}
\label{tab:tablelens}
\end{table*}

If we combine CMB with background data from SNIa, BAO and cosmic chronometers, we get even tighter constraints. The values of $d$ are still $\lesssim 1\sigma$ away from 1 and $-2.6\lesssim \Delta{\rm DIC}\lesssim +0.5$, so again we find no hint of new physics, e.g. we get $d=0.987^{+0.025}_{-0.028}$ with the Planck18(lens)+SNIa+BAO+CCH dataset (cf. Table \ref{tab:tablelens}). Volume effects are not expected to have a significant impact on these results \cite{AGV2022}.

The inclusion of the SH0ES data makes a difference, as it is evident from the very large positive values of $\Delta{\rm DIC}$ obtained when CMB is combined with background data and the measurements from SH0ES. They read $\Delta{\rm DIC}\gtrsim +17$ regardless of the concrete CMB data configuration employed in the fitting analysis. There is an important lowering of the value of $d$, which departs now from 1 at $\gtrsim 4.5\sigma$ C.L. We obtain $d=0.924\pm0.017$ with Planck18+SNIa\_$H_0$+BAO+CCH. This leads to $G=(1.082\pm 0.020)G_N$, so to a central value of $G$ $\sim 8\%$ larger than $G_N$. This is aligned with previous results in the literature found in the context of Brans-Dicke with a cosmological constant \cite{BDLCDM_CQG,BD-RVM} and the type-II RRVM \cite{rvmtens,rvmtens2}, which exhibit a similar preference for larger values of $G$ when the SH0ES information is considered on top of CMB, background and LSS data in the fitting analyses. See also \cite{Benevento2022}. This clear departure of the value of $G$ preferred at cosmological scales from the one measured locally can be thought of as natural in the context of models with a space-time dependent gravitational coupling, but it requires a physical mechanism that explains the  transition between the cosmological and astrophysical regimes, see e.g. \cite{Amendola:2015ksp,Clifton:2011jh,Avilez:2013dxa,Gomez-Valent:2021joz}. This is beyond the scope of this paper, of course, but in any case, we expect this mechanism to enter into play at non-linear scales and, hence, at scales that have not been proved by our dataset. The lowering of $d$ is of course accompanied by non-negligible ($\sim 2\sigma$) shifts in essentially all the parameters of the model, including those that are relevant for the discussion of the cosmological tensions. A comparison of the constraints on the relevant parameters of the StRVM obtained with Planck18 alone with those found including also the background and SH0ES information is provided in Fig. \ref{fig:contours}. The improvement of the StRVM versus the concordance $\Lambda$CDM model (whose contours are also included in that figure) is rendered evident on simple inspection. In particular, the $H_0$ tension is virtually absent as it is below the $1\sigma$ C.L. -- if estimated through the comparison of the SH0ES measurement of $H_0$  ($H_0=73.04\pm1.04$ km/s/Mpc) and our fit values in the StRVM. If, however, we use our posterior values for the absolute magnitude of SNIa and compare them to the SH0ES determination ($M=-19.253\pm 0.027$ mag) we are led  to a moderate discrepancy of $\sim 2\sigma$ C.L. We find, in any case, a notable decrease of the tension with SH0ES in the context of the StRVM. Worth noticing in Fig. \ref{fig:contours} is also the shift of $\sigma_{12}$  in the right direction, viz. towards smaller values than those predicted in the $\Lambda$CDM model. The impact on $S_8$, instead, is not so important, but according to Ref. \cite{Sanchez2020} the most adequate parameter for judging the growth tension is $S_{12}$. Hence, on these grounds, it seems that the physical mechanism that allows to alleviate the $H_0$ tension also produces a lower amplitude of the matter power spectrum at linear scales, which is in agreement with what we have explained in Sec. \ref{sec:perturbations} (see the caption of Fig. \ref{fig:Pk_d}). This is remarkable. However, the fit prefers to accommodate well the SH0ES data at the expense of worsening the description of the CMB. This can be also seen in the tables of Appendix \ref{sec:extraTables}, specially in Tables \ref{tab:table_chi2_I} and \ref{tab:table_chi2_III}, which include the contribution of the CMB polarization data. As discussed in Sec. \ref{sec:perturbations}, the polarization data partially breaks the degeneracy between $n_s$ and $d$. If we remove the polarizations, i.e. if we consider Planck18(np) in the fitting analyses, the situation is much more favorable for the StRVM, since in this case the inclusion of SH0ES does not induce a significant raise of the $\chi^2_{\rm CMB}$ (cf. Table \ref{tab:table_chi2_II}).

\begin{table*}[t!]
\centering
\begin{tabular}{|c ||c | c ||c || c|c| }
 \multicolumn{1}{c}{} & \multicolumn{1}{c}{}\\
\multicolumn{1}{c}{} &   \multicolumn{5}{c}{Planck18+SNIa\_$H_0$+BAO+CCH}
\\\hline
{\small Parameter} & {\small $\Lambda$CDM $(\alpha_s)$}  & {\small StRVM $(\alpha_s)$} & {\small StRVM $(\beta_s)$} & {\small $\Lambda$CDM $(A_L)$}  & {\small StRVM $(A_L)$}
\\\hline
$10^2\omega_b$ &  $2.262\pm 0.014$ & $2.277\pm 0.014$ & $2.279\pm 0.014$  & $2.272\pm 0.014$  & $2.293\pm 0.015$  \\\hline
$\omega_{\rm cdm}$ & $0.1170\pm 0.0008$ & $0.1180\pm 0.0008$  &$0.1181\pm 0.0008$ & $0.1162\pm 0.0008$ & $0.1173\pm 0.0008$  \\\hline
$n_s$ & $0.968\pm 0.004$ & $0.983\pm 0.005$ & $0.980\pm 0.005$ & $0.973\pm 0.004$ & $0.984\pm 0.005$   \\\hline
$\tau_{\rm reio}$ & $0.060^{+0.008}_{-0.009}$ & $0.058^{+0.008}_{-0.009}$& $0.060^{+0.008}_{-0.010}$ & $0.050^{+0.009}_{-0.008}$  &$0.051\pm 0.008$\\\hline
$\sigma_{12}$ & $0.788^{+0.008}_{-0.009}$ & $0.778\pm 0.015$ & $0.780\pm 0.015$ & $0.776\pm 0.009$ & $0.766\pm 0.015$  \\\hline
$H_0$ & $68.42^{+0.33}_{-0.34}$ & $71.39^{+0.79}_{-0.82}$ & $71.21^{+0.79}_{-0.77}$ &$68.80^{+0.36}_{-0.37}$& $71.54^{+0.76}_{-0.78}$\\\hline
$d$ & $-$ & $0.921^{+0.017}_{-0.019}$ & $0.925^{+0.016}_{-0.017}$ & -& $0.927\pm 0.017$
\\\hline
$\alpha_s$ & $-0.0060\pm0.0070$ & $0.0050^{+0.0073}_{-0.0074}$ & $-$ & $-$ & $-$
\\\hline\hline
$\beta_s$ & $-$ & $-$ & $0.0023^{+0.0087}_{-0.0093}$ & $-$ & $-$
\\\hline\hline
$A_L$ & $-$ & $-$ & $-$ & $1.219^{+0.060}_{-0.064}$ & $1.211^{+0.064}_{-0.062}$
\\\hline\hline
$r_d$ & $147.79\pm 0.23$ & $141.30^{+1.47}_{-1.56}$ & $141.57^{+1.40}_{-1.46}$ & $147.87^{+0.21}_{-0.22}$ & $141.75\pm 1.44$ \\\hline
$M$ & $-19.410\pm 0.010$ & $-19.319^{+0.023}_{-0.024}$ & $-19.324^{+0.024}_{-0.023}$  & $-19.400^{+0.010}_{-0.011}$ & $-19.316\pm 0.023$ \\\hline
$S_{8}$ & $0.800\pm 0.011$ & $0.817\pm 0.018$& $0.818^{+0.018}_{-0.0017}$ & $0.784\pm 0.011$  & $0.799^{+0.017}_{-0.019}$ \\\hline
$\sigma_8$ & $0.802^{+0.007}_{-0.008}$ & $0.816^{+0.017}_{-0.016}$ & $0.817^{+0.017}_{-0.016}$  & $0.793^{+0.008}_{-0.007}$  & $0.804^{+0.016}_{-0.017}$
 \\\hline
$S_{12}$ & $0.787\pm 0.009$ & $0.780\pm 0.015$ & $0.782^{+0.016}_{-0.015}$  & $0.773\pm 0.010$  & $0.767^{+0.015}_{-0.016}$ \\\hline
$\chi^2_{\rm min}$ & 4292.96  &   4272.84 & 4273.54 & 4279.48  & 4262.84 
\\\hline
$\Delta{\rm DIC}$ & -0.73  &  +16.75 & +16.25 & +11.26 & +29.18 
\\\hline
\end{tabular}
\caption{Fitting results obtained with the Planck18+SNIa\_${H_0}$+BAO+CCH dataset and allowing the following parameters to vary freely in the Monte Carlo: (i) the running of the spectral index, $\alpha_s$; (ii) the running of the running, $\beta_s$; (iii) the $A_L$ parameter. In the last row we show the DIC differences computed using in all cases as reference the $\Lambda$CDM model with $\alpha_s=\beta_s=0$ and $A_L=1$, {\it cf.} the penultimate column of Table \ref{tab:tableI}. The values of $\chi^2_{\rm min}$ obtained with a running spectral index are almost identical to those provided in the last two columns of that table. This is why $\Delta$DIC are slightly lower, since here we have one additional parameter ($\alpha_s$ or $\beta_s$). We can conclude that the running of $n_s$ does not help significantly to improve the description of the data. We get values of $\alpha_s$ and $\beta_s$ which are compatible with 0 at $\lesssim 1\sigma$ C.L. On the other hand, we find a $\gtrsim 3\sigma$ preference for $A_L>1$ and a prominent increase of $\Delta$DIC in the $\Lambda$CDM ($A_L$) and StRVM ($A_L$) with respect to the scenarios with $A_L=1$. StRVM cannot alleviate significantly the CMB lensing anomaly.}
\label{tab:tableII}
\end{table*}

We have also studied whether the aforementioned issue concerning the polarization data can be mitigated by considering a running spectral index in the primordial power spectrum. We present the corresponding results in Table \ref{tab:tableII}. It turns out that such an increase of the complexity of the model does not produce a significant decrease of the $\chi^2_{\rm min}$ compared to the value obtained with a rigid (non-running) $n(k)=n_s$, see the last column of Table \ref{tab:tableI}. According to the deviance information criterion, the addition of these extra degrees of freedom is not justified.

Finally, we have also studied the status of the lensing anomaly in the context of the StRVM. We do so by allowing the CMB lensing consistency parameter $A_L$ to vary freely in the Monte Carlo. The results are also presented in Table \ref{tab:tableII}. We still find values of $A_L$ larger than one at $\gtrsim 3\sigma$ C.L. and, therefore, a very similar level of tension to the one found in the $\Lambda$CDM.


 \begin{figure}[t!]
\begin{center}
 \includegraphics[width=6.2in, height=5.5in]{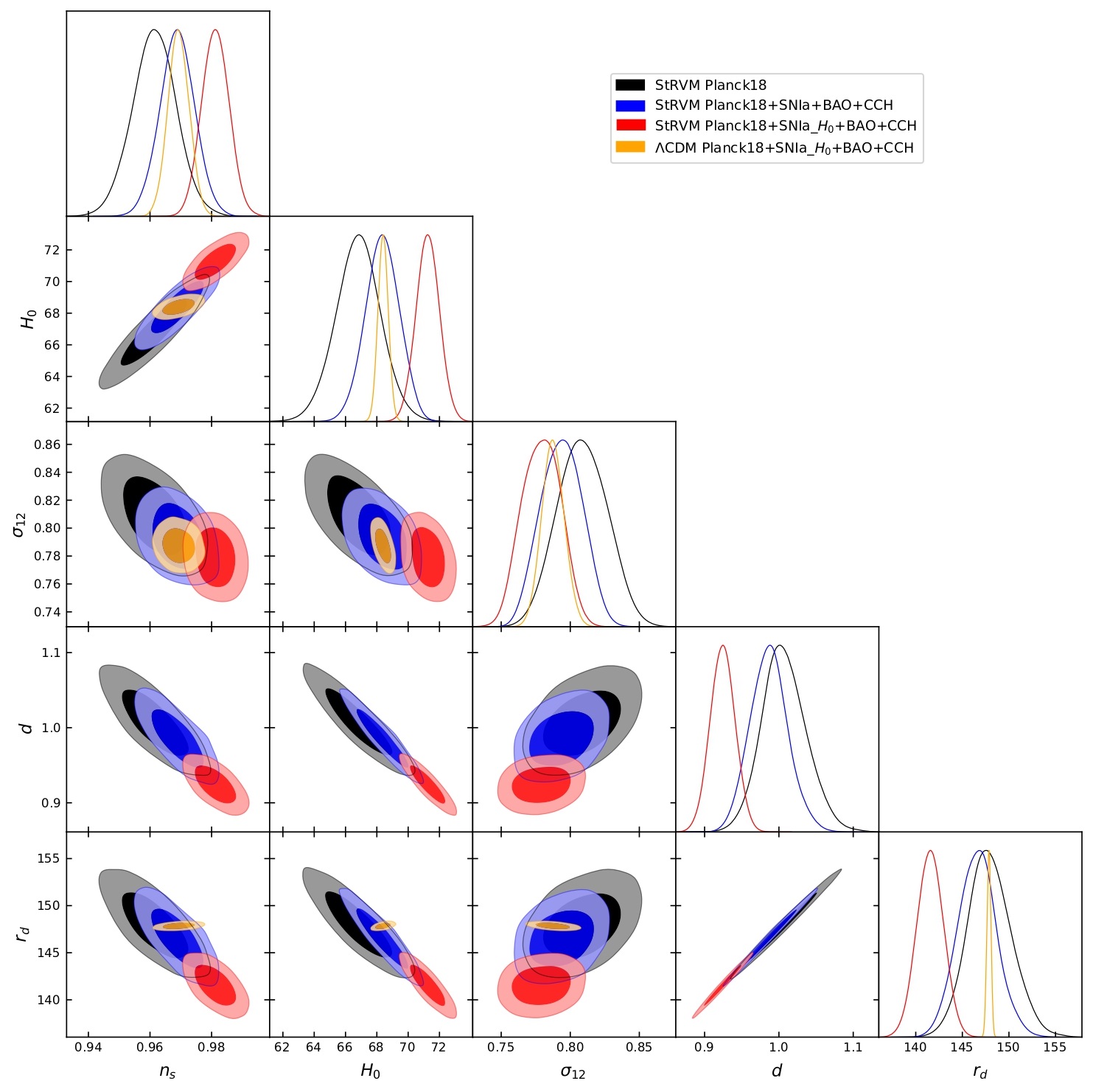}
 \caption{Contour plots and one-dimensional posterior distributions for the parameters $(H_0,\sigma_{12},d,r_d,n_s)$ of the StRVM obtained with the datasets Planck18, Planck18+SNIa+BAO+CCH and Planck18+SNIa\_$H_0$+BAO+CCH. For the latter dataset, we also show the results obtained with $\Lambda$CDM.}\label{fig:contours}
\end{center}
\end{figure}

\section{Energy Conditions and TCC in the StRVM}\label{sec:ectcc}

Before closing, we feel like discussing briefly, for completion, two issues, whose resolution though falls beyond the scope of the current article. The first concerns the validity of the energy conditions in our effective theory \eqref{eq:action} in all the post-inflationary epochs. As well known~\cite{Visser}, it is expected that at least at some point in the evolution of the Universe, the energy conditions would be violated. These conditions characterise dark energy models. Whether these conditions are satisfied for the entire history of our string-inspired model is not essential for the analysis in this work, given that, as discussed in \cite{ms1,ms2}, the nature of the current-era dark energy in the StRVM is still not understood from a microscopic point of view~\cite{bms1}, and moreover, as we shall discuss below, it depends crucially on details of the underlying string model.
In general, like in other cosmology models, we can expect a change in the status of the energy conditions at different cosmic periods. As a matter of fact, there is nothing sacrosanct about the energy conditions that the DE must obey. This concept remains unaccounted for on fundamental grounds.  At present there is a wide range of energy conditions used to describe the DE and nothing forbids a transition between them, 
provided the expansion properties of the universe are in accordance with observations. Among the various possibilities, 
 we mention, for concreteness, two popular (and very disparate)  examples, namely, 
quintessence (which violates the strong energy condition but preserves the weak and dominant energy conditions -- see below),  and phantom dark energy, which in fact  violates  the entire set of classical energy conditions!  Many other exotic possibilities can occur, see {\it e.g.} Fig. 3 of \cite{ms2} and discussion therein.

Thus, in view of the diversity of situations for the description of the DE, it would be interesting to briefly study the status of the energy conditions  within the context of our effective point-like theory \eqref{eq:action}. We start discussing the Null Energy Condition (NEC), which takes the following covariant form: 

\begin{equation}
T_{\mu\nu}k^\mu k^\nu\geq 0\,,
\end{equation}
for any null vector ($k^\mu k_\mu=0$). In a FLRW universe, the NEC can be written explicitly as follows if we consider the energy-momentum tensor of the total cosmological fluid:

\begin{equation}\label{eq:NEC0}
\sum_{i=m,r,{\rm vac}}(\rho_i+p_i)\geq 0\,,
\end{equation}
or, equivalently,

\begin{equation}
R_{\mu\nu}k^\mu k^\nu\geq 0\longrightarrow \dot{H}\leq 0\,.
\end{equation}
We can use Eqs. \eqref{eq:Fr1} and \eqref{eq:Fr2} of our paper to write inequality \eqref{eq:NEC0} in terms of the matter and radiation energy densities and pressures, and the parameter $\epsilon$. Taking into account that $G(a)>0\,\forall{a}$ for the typical values of $\epsilon$ allowed by the data we find, 

\begin{equation}\label{eq:NEC}
\sum_{i=m,r}(\rho_i+p_i)+\frac{\epsilon}{\left(1+\frac{4c_0}{\rho_m}\right)^2}\left[3\rho_r+c_0-\frac{5}{2}\rho_m+\frac{2c_0}{\rho_m}(8c_0+11\rho_m+12\rho_r)\right]\geq 0\,.
\end{equation}
If $\epsilon=0$ we recover the NEC condition in $\Lambda$CDM, i.e. $\sum_{i=m,r}(\rho_i+p_i)\geq 0$ (in which the cosmological term $\Lambda$ yields zero contribution). From inequality \eqref{eq:NEC} one can see that the total cosmological fluid in StRVM fulfills the NEC in all the stages of the cosmic expansion provided $|\epsilon|\ll 1$. This is in fact the case in our analysis, see Eq.\,\eqref{eps7}. The effective vacuum, though, violates the NEC if $\epsilon<0$.

The Strong Energy condition (SEC) imposes the NEC and 

\begin{equation}
\left(T_{\mu\nu}-\frac{1}{2} T^\alpha_\alpha g_{\mu\nu}\right)u^\mu u^\nu\geq 0\,,
\end{equation}
where $u$ is a timelike four-vector ($u^\mu u_\mu>0$). If we consider the total energy-momentum tensor, this inequality reduces to $R_{\mu\nu}u^\mu u^\nu\geq 0$, which can be expressed in a very simple way in a FLRW universe: 
\begin{equation}
q\geq 0\,,
\end{equation}
with $q=-\ddot{a}a/\dot{a}^2$ the deceleration parameter. This tells us that the SEC is only satisfied by the total cosmological fluid in a decelerating universe. This is also equivalent to

\begin{equation}
\sum_{i=m,r,{\rm vac}}(\rho_i+3p_i)\geq 0\,,
\end{equation}
where in our case the explicit form of the vacuum energy density and pressure can be obtained again from Eqs. \eqref{eq:Fr1} and \eqref{eq:Fr2}. It is easy to convince oneself that the SEC is violated at late times in StRVM, when there is acceleration. This is actually analogous to what happens in $\Lambda$CDM. The SEC is violated by the effective vacuum if $\epsilon<0$ since, as discussed above, it also violates the NEC. If, instead, $\epsilon>0$, the effective vacuum violates the SEC at late times and, depending on the exact value of $\epsilon$, there might be also a violation in some periods of the cosmic expansion, at higher redshifts.

The Weak Energy condition (WEC), on the other hand, considers the NEC together with 

\begin{equation}\label{eq:WEC}
T_{\mu\nu}u^\mu u^\nu\geq 0\,.
\end{equation}
Let us start by studying the WEC applied to the total cosmological fluid, which incorporates the contributions of radiation, matter and the effective vacuum fluid described by the energy-momentum tensor \eqref{eq:Tvac2}. If we treat these components as perfect fluids, inequality \eqref{eq:WEC} can be written as follows,

\begin{equation}
\sum_{i=m,r,{\rm vac}}\rho_i\geq 0\,.
\end{equation}
In StRVM, the WEC is satisfied by the total effective cosmological fluid, regardless of the post-inflationary epoch under consideration. We can also study whether the energy-momentum tensor associated to the vacuum, as defined in Eq. (32) of our paper, fulfills the WEC. In this case, the condition takes the form 

$$\rho_{\rm vac}\geq 0\,,$$

or, equivalently, 

\begin{equation}
c_0+\frac{\epsilon}{2}(4c_0+\rho_m)+\frac{3\epsilon}{1+\frac{4c_0}{\rho_m}}(\rho_m+\rho_r+c_0)\geq 0\,.
\end{equation}
Thus, at sufficiently early stages of the cosmic history the WEC is violated by the vacuum if $\epsilon<0$, whereas it is always fulfilled for positive values of $\epsilon$.

Finally, we discuss the Dominant Energy condition (DEC), which assumes the WEC and, in addition, 

\begin{equation}
(T_{\mu\nu}V^\nu)(T^{\mu\alpha}V_\alpha)\geq 0\,,
\end{equation}
for any time-like vector ($V_\mu V^\mu>0$). If applied to the total cosmological fluid, it can be expressed as follows, 

\begin{equation}
\sum_{i=r,m,{\rm vac}}\rho_i\geq |\sum_{i=r,m,{\rm vac}}p_i|\,.
\end{equation}
This condition is satisfied during the radiation-, matter- and vacuum-dominated epochs. If we focus only on the vacuum, the DEC simply reads

\begin{equation}\label{eq:DECvac}
\rho_{\rm vac}\geq |p_{\rm vac}|\,.
\end{equation}
Our effective vacuum does not necessarily satisfy the DEC at sufficiently high redshifts because, as mentioned before, it may violate the WEC (when $\epsilon<0$). The DEC, though, will be respected in our framework for $\epsilon >0$ when the vacuum  eventually dominates over matter and radiation in the remote future. This can be seen upon using again Eqs. \eqref{eq:Fr1} and \eqref{eq:Fr2} in the inequality \eqref{eq:DECvac} and the fact that the condition $\rho_m, \rho_r \ll c_0$ will become 
$\epsilon\, \rho_m \Big( 1 + \mathcal O(\rho_m/c_0)\Big) \ge 0$, since $p_{\rm vac} < 0$,
which 
will eventually be fulfilled for $\epsilon > 0$, in the future cosmic history, in which case an approximate RVM de-Sitter equation of state (saturating \eqref{eq:DECvac}) would arise. For negative $\epsilon$, on the other hand, there will be a very small (negligible for all practical purposes) violation of DEC in the far future.

Nonetheless, in the context of a string theory model, 
such a point-like effective field theory analysis is 
not sufficient. When the model \eqref{eq:action} is viewed as a low-energy limit of some underlying string theory model after appropriate compactification, the validity of the energy conditions in the four-dimensional space-time model becomes more complicated, depending on details of compactification.  In our effective approach we did not discuss such details. Those can lead to
stringent constraints when comes to the question as to whether the energy conditions are satisfied. Indeed, as discussed in \cite{ec1,ec2}, in models in which a NEC is satisfied, a dark energy phase of the compactified-string-inspired universe, consistent with observations, 
is only possible at late eras, if both Newton’s gravitational constant and the dark energy equation-of-state
vary with time. 

Such features seem to characterise our effective theory, but it goes without saying that a detailed embedding of our theory on specific string theories and specific compactification schemes fall beyond the scope of the present article. Having said that, though, we also remark that our string inspired cosmology at late epochs, which has been the focus of our work in this article, has its origin in supergravity theories which describe low-energy effective string theories at early times. In \cite{ec3}, it has been argued that higher-dimensional string-compatible supergravity actions, corresponding to the 
standard content of the massless string spectrum, as in our case in this article, do satisfy all the energy conditions, provided quantum corrections are ignored. Such higher-dimensional energy conditions might induce the preservation of the four-dimensional energy conditions, so we might expect this to characterise our models.

Another issue that we have not discussed in detail is the satisfaction (or not) by our effective theory of the TCC~\cite{tcc1}, which is known to follow~\cite{tcc2,tcc2b} from the swampland distance conjecture~\cite{swamp4}, and, more generally,  to have implications for the various swampland conjectures~\cite{tcc3}.
The TCC implies the important restriction that modes that cross the Hubble horizon could ever have had a wavelength smaller
than the Planck length (or, equivalently, their momentum is less than the Planck energy scale $M_{\rm Pl}$).

In our theory~\cite{ms1,ms2}, as we have stressed above and in our previous articles, transplanckian modes are not allowed, as the ultraviolet momentum cutoff is set to the string scale $M_s < M_{\rm Pl}$. However, the TCC, as formulated in \cite{tcc1}, implies that, in models of standard inflation, induced by an inflaton field, the energy scale of inflation is at most $10^{9}$~GeV, which would imply extremely small slow-roll parameters\footnote{Here we denote the inflationary slow-roll parameter as $\epsilon_I$ (instead of $\epsilon$) to distinguish it from the parameter of the StRVM \eqref{epsdef}.} $\epsilon_I < 10^{-31}$, so that the amplitudes of cosmological fluctuations agree with observations, which would imply an extremely small  tensor to scalar ratio $< 10^{-30}$.

As we have discussed in detail in \cite{bms1,bms2,ms1,ms2}, our string-inspired cosmology is characterised by an inflationary phase that is induced not by external inflaton fields, but is due to the RVM-like non linearities that are induced by condensates of the anomalous gravitational Chern-Simons terms. In our case, the slow roll parameter is provided by the axion-like field that is associated with the four space-time dimensional dual of the Kalb-Ramond field strength. This evades the TCC argumens of \cite{tcc1}, and, as shown in \cite{ms1,ms2}, the corresponding inflationary epochs are consistent with large inflationary scale $H_I \sim 10^{-5}\, M_{\rm Pl}$, in agreement with observations~\cite{planck}.

Nonetheless, it should be remarked that the considerations and estimates of \cite{ms1,ms2} 
on the condensates of the anomalous gravitational Chern-Simons terms, which leads to inflation, are based on effective local field theories. A full derivation of the quantum-gravity-induced anomalous Chern-Simons condensates requires a complete understanding 
of the r\^ole of the infinite towers of massive stringy states, whose manipulation is at present not feasible. Thus, from the point of view of a full string theory, we consider the issue as wide open.

\section{Summary, discussion and outlook}\label{sec:concl} 

 In this work we have used the framework of a string-inspired cosmology of RVM type, developed in \cite{bms1,bms2,ms1,ms2}, to discuss its modern-day phenomenology and in particular its ability to alleviate {\it both} the Hubble-$H_0$ tension and the tension in the galactic growth data. An important ingredient of such models is the inclusion of quantum-gravity fluctuations, which although generated at primordial epochs, when the model is embedded in appropriate dynamically-broken  supergravity theories,
nonetheless they lead to corrections to the vacuum energy density which exhibit logarithmic dependence on the Hubble parameter in modern times of the form  $\tilde c_2\, H^2 {\rm ln}(M_{\rm pl}^2/H^2)$. The coefficient $\tilde c_2$ is proportional to the scale of the bare cosmological constant $\kappa^2 \, |\mathcal E_0 | $
or $\kappa^2 \mathcal E_0 \, {\rm ln}(\kappa^4 |\mathcal E_0|)$, {\it cf.} \eqref{cs}. In the case of the supergravity model, $|\mathcal E_0| = f^2$, where $\sqrt{f}$ is the energy scale of the primordial dynamical supergravity breaking. It should be remarked that in such scenarios, the logarithmic-$H$ quantum-graviton-induced corrections are of primordial origin, surviving until today. 

Another closely related kind of logarithmic corrections that can survive in our present universe are those of the type  $\sim H^2 {\rm ln}(m^2/H^2)$,  see Eqs.\,\eqref{qft}-\eqref{nueffz}, which appear when we take into account the quantum effects generated by the fluctuations of the quantized matter fields in curved spacetime, which can be significant for masses of order of a GUT scale, $m\sim M_X$\,\cite{Moreno-Pulido:2020anb,Moreno-Pulido:2022phq,Moreno-Pulido:2023ryo}. However, in a scenario where $\sqrt{f}$ is sufficiently close to the reduced Planck energy scale, the quantum gravity corrections may dominate over the corresponding ones from  integrating out quantum  matter fields \eqref{nueffz}. In this paper, we have precisely focused on exploiting this possibility, but in general these two types of effects could actually be present and behave collaboratively in the task of smoothing out the cosmological tensions under study.  We have put very strong constraints on the running of the vacuum arising from graviton effects in the context of the StRVM. In order not to spoil the correct description of the large-scale structure at linear scales we have to demand a subplanckian $\sqrt{f}$ close to the reduced Planck scale \eqref{f7}, consistent with the dynamical supergravity scenario of \cite{ms1,houston1}.  

On the other hand, for generic non-supersymmetric quantum gravity models, such logarithmic $\tilde c_2 \, H^2 \, {\rm ln}(\kappa^2 H^2)$ corrections to the vacuum energy density also exist, but in such cases $|\mathcal E_0|$
is a bare cosmological constant which is a free phenomenological parameter. In such models, the one-loop renormalised (``R'') cosmological constant is the one which is identified with the observed one $|\mathcal E_{\rm R} | \sim 10^{-122}\, \kappa^{-4}$, and not $|\mathcal E_0|$. If $\mathcal E_0 \sim |\mathcal E_{\rm R}|$, then such terms may be subleading  compared to the quantum-matter-induced terms \eqref{nueffz}.

Our phenomenological analysis in this work has also demonstrated that, 
if the parameter $d$ \eqref{eq:ddeffG}, which defines the ratio of Newton's constant to the current value of the gravitational coupling acting at cosmological scales -- as follows from the graviton equations of motion stemming from \eqref{c1c2} and the effective action 
\eqref{eq:action} --, takes on values in the range $d \sim 0.90-0.95$, then the model leads to an alleviation of both tensions, the $H_0$ and the growth tension. We have assessed the situation of the growth data both through the traditional $\sigma_8$ parameter and also in terms of the alternative one $\sigma_{12}$, which has been claimed to be a more realistic quantity for the analysis of the growth tension -- see \cite{{Sanchez2020}}. We remark that the aforesaid preferred  low values of the parameter $d$ below $1$ (or, equivalently, enhanced values of $G(z=0)$ $\gtrsim 8\%$ larger than $G_N$)  are obtained only when the SH0ES data are also considered in our fitting analyses. The signal for new physics reaches the $\sim 4.5$ C.L. when we include the CMB polarization anisotropies, but can grow up to the $5.2\sigma$ C.L. when we do not consider the high-$\ell$ polarization data. These results are also strongly supported by Bayesian information criteria. We remind the reader at this point that the clustering of matter depends on whether $d$ is larger or less than one, as discussed in section \ref{sec:perturbations}. Indeed, for $d < 1$ and fixed values of the other parameters there is a decrease of the power spectrum, in contrast to the $d \gtrsim 1$ case (cf. Fig. \ref{fig:Pk_d}).

It should be stressed, at this point, that the effective gravitational constant $G(z)$ receives corrections from {\it both} matter and quantum graviton contributions. Thus constraining phenomenologically the value of the parameter $d$ and the running of the vacuum, e.g. by requiring the alleviation of the cosmological tensions, might provide important insight on the underlying microscopic theory. The link between the cosmological and local regimes is, however, not trivial at all. In particular, a transition from the cosmological $G$ to $G_N$ at astrophysical domains is required to explain the best-fit values needed to loosen the tensions, but this study is certainly beyond the scope of this paper.


\acknowledgments
The work  of AGV is funded by the Istituto Nazionale di
Fisica Nucleare (INFN) through the project of the InDark INFN Special Initiative: ``Dark Energy and Modified Gravity Models in the light of Low-Redshift Observations'' (n. 22425/2020). That of 
NEM is supported partly by STFC (UK) Grants ST/T000759/1 and ST/X000753/1, while the work of JSP is funded in part by the projects PID2019-105614GB-
C21, FPA2016-76005-C2-1-P (MINECO, Spain),  2021-SGR-249 (Generalitat de Catalunya) and
CEX2019-000918-M (ICCUB, Barcelona). AGV and JSP take part in the COST Association Action CA21136 ``{\it Addressing observational tensions in cosmology
with systematics and fundamental physics (CosmoVerse)}''.
NEM and JSP also acknowledge participation in the COST Action CA18108 ``{\it Quantum Gravity Phenomenology in the Multimessenger Approach (QG-MM)}''. 
\color{black}


\appendix

\section{Cosmological perturbations in the synchronous gauge}\label{sec:appendixPert}

\noindent In this appendix we provide the expressions of the geometrical quantities and the energy-momentum tensors of the cosmological fluids and the vacuum up to linear order in the perturbed quantities. We use the synchronous gauge, since this is the one employed to solve the set of coupled differential equations numerically in our modified version of the Einstein-Boltzmann solver \texttt{CLASS} \cite{CLASS1,CLASS2}. We use, as in the main text, the sign convention $(-,+,+)$ in the classification by Misner {\it et al.} \cite{Misner}. In the synchronous gauge the line element reads,

\begin{equation}
ds^2 = a^2(\tau) [d\tau^2-(\delta_{ij}+h_{ij}(\tau,\vec{x}))\,dx^idx^j]\,,
\label{confmetr}
\end{equation}
with $\tau$ the conformal time, $\vec{x}$ the spatial comoving coordinates, and $h_{ij}$ the metric perturbation. The components of the metric tensor and its inverse take the following form, respectively,

\begin{equation}
g_{00}=a^2\qquad g_{ij}=-a^2(\delta_{ij}+h_{ij})\quad ||\quad g^{00}=a^{-2}\qquad g^{ij}=-a^{-2}(\delta_{ij}-h_{ij})\,.\\
\end{equation}

\subsection{Geometrical quantities}

In the synchronous gauge, the perturbed FLRW metric has the following associated non-null elements of the Christoffel symbols,

\begin{equation}\label{eq:Christoffel}
    \begin{split}
        \Gamma^{0}_{00}=&\mathcal{H}\quad;\quad \Gamma^{0}_{i0}=0\quad;\quad \Gamma^{0}_{ij}=\mathcal{H}(\delta_{ij}+h_{ij})+\frac{h^\prime_{ij}}{2}\\
        \Gamma^{i}_{00}=&0\quad ;\quad \Gamma^{i}_{j0}=\mathcal{H}\delta_{ij}+\frac{h^\prime_{ij}}{2}\quad;\quad \Gamma^{i}_{jl}=\frac{1}{2}(h_{ij,l}+h_{il,j}-h_{jl,i})\,,
    \end{split}
\end{equation}
with the primes denoting derivatives with respect to the conformal time, the lower comas partial derivatives with respect to the spatial comoving coordinates, and $\mathcal{H}\equiv a^\prime/a$. Using \eqref{eq:Christoffel} one can compute the components of the Ricci tensor, 

\begin{equation}
    \begin{split}
        R_{00}=&-3\mathcal{H}^\prime-\frac{h^{\prime\prime}}{2}-\frac{\mathcal{H}}{2}h^\prime\,,\qquad  R_{0i} = \frac{\partial_j h^\prime_{ij}}{2}-\frac{\partial_i h^\prime}{2}\,,\\
        R_{ij}=&(\delta_{ij}+h_{ij})(\mathcal{H}^\prime+2\mathcal{H}^2)+\frac{h^{\prime\prime}_{ij}}{2}+\frac{1}{2}\left(h_{li,jl}+h_{lj,il}-h_{ij,ll}-h_{,ij}\right)+\frac{\mathcal{H}}{2}h^\prime\delta_{ij}+\mathcal{H}h^\prime_{ij}\,,
    \end{split}
\end{equation}
and, hence, also the Ricci scalar, 

\begin{equation}
a^2R=-6(\mathcal{H}^\prime+\mathcal{H}^2)-h^{\prime\prime}-3\mathcal{H}h^\prime-h_{li,li}+h_{,ll}\,.
\end{equation}
These results can be used to finally compute the elements of the Einstein tensor, 

\begin{equation}\label{eq:PerturbEinstTensor}
\begin{split}
G_{00}=&3\mathcal{H}^2+\mathcal{H}h^\prime+\frac{1}{2}\left(h_{li,li}-h_{,ll}\right)\,,\\
G_{i0}=&\frac{1}{2}\left(\partial_jh^\prime_{ij}-\partial_{i}h^\prime\right)\,,\\
G_{ij}=&-(\delta_{ij}+h_{ij})(2\mathcal{H}^\prime+\mathcal{H}^2)+\frac{h^{\prime\prime}_{ij}}{2}-\frac{h^{\prime\prime}}{2}\delta_{ij}-\mathcal{H}h^\prime\delta_{ij}+\mathcal{H}h^\prime_{ij}\\
&+\frac{1}{2}(h_{li,jl}+h_{lj,il}-h_{ij,ll}-h_{,ij}-h_{lt,lt}\delta_{ij}+h_{,ll}\delta_{ij})\,.
\end{split}
\end{equation}
The scalar part of the metric perturbation can be written as follows \cite{MaBertschinger},     

\begin{equation}         
h_{ij}(\tau,\vec{x}) = \int d^3k\, e^{-i\vec{k}\cdot \vec{x}}\left[h(\tau,\vec{k})\,\hat{k}_i\hat{k}_j+6\,\eta(\tau,\vec{k})\left(\hat{k}_i\hat{k}_j-\frac{\delta_{ij}}{3}\right)\right]\,,
\end{equation}
with $\hat{k}=\vec{k}/k$ and $k=|\vec{k}|$. $h(\tau,\vec{k})$ is the trace of $h_{ij}$ in momentum space, whereas $\eta(\tau,\vec{k})$ is the function that controls its traceless part. It is useful to write the perturbed components of the Ricci scalar and the Einstein tensor in momentum-space, since they will be employed later on in Einstein's equations. They read, respectively,

\begin{equation}
    \delta R= a^{-2}(4\eta k^2-3\mathcal{H}h^\prime-h^{\prime\prime})\,,
\end{equation}
and

\begin{equation}\label{eq:PerturbEinstTensor2}
\begin{split}
\delta G_{00}=&\mathcal{H}h^\prime-2\eta k^2\,,\\
\delta G_{i0}=&-ik_i2\eta^\prime\,,\\
\delta G_{ij}=&\hat{k}_i\hat{k}_j\left[-(2\mathcal{H}^\prime+\mathcal{H}^2)(h+6\eta)+\frac{h^{\prime\prime}+6\eta^{\prime\prime}}{2}+\mathcal{H}(h^\prime+6\eta^\prime)-\eta k^2\right]\\
&+\delta_{ij}\left[2\eta(2\mathcal{H}^\prime+\mathcal{H}^2)-\eta^{\prime\prime}-\frac{h^{\prime\prime}}{2}-\mathcal{H}h^\prime-2\eta^\prime\mathcal{H}+\eta k^2\right]\,.
\end{split}
\end{equation}


\subsection{Energy-momentum tensor of a perfect fluid}

\noindent The energy-momentum tensor of a perfect fluid reads, 

\begin{equation}
T_{\mu\nu}=(p+\rho)u_\mu u_\nu-pg_{\mu\nu}\,,
\end{equation}
The perturbed 4-velocity is

\begin{equation}
u^\mu = \frac{1}{a}(1,v^i)\qquad ;\qquad u_\mu = a(1,-v^i)\,.
\end{equation}
We split the energy density and pressure as $\rho=\bar{\rho}+\delta\rho$ and $p=\bar{p}+\delta p$, respectively, with the bar denoting the background contribution. Hence, at linear order we find,  

\begin{equation}
T_{00}= a^2(\bar{\rho}+\delta\rho)\,, \quad T_{0i} = -a^2(\bar{\rho}+\bar{p})v^i\,,\quad T_{ij}=a^2\delta_{ij}(\bar{p}+\delta p)+a^2 h_{ij}\bar{p}\,.
\end{equation}
Some species, as neutrinos, can receive also a contribution from anisotropic shear, see {\it e.g.} \cite{MaBertschinger}. We duly take this fact into account in our numerical code. 


\subsection{Energy-momentum tensor of the vacuum}

The perturbation of the vacuum EMT \eqref{eq:Tvac2} is given by the following expression, 

\begin{equation}
    \begin{split}
        \delta T^{\rm (vac)}_{\mu\nu} =\, & \delta g_{\mu\nu}\left(c_0-\frac{\nu \bar{R}}{16\pi G_N}\right)-\bar{g}_{\mu\nu}\frac{\nu\,\delta R}{16\pi G_N} +\frac{\nu}{8\pi G_N}\left[(\delta^{\alpha}_\mu\delta^\beta_\nu-\bar{g}_{\mu\nu}\bar{g}^{\alpha\beta})\left(\frac{\partial_\beta\partial_\alpha \delta R}{\bar{R}}-\frac{\delta R}{\bar{R}^2}\partial_\beta\partial_\alpha \bar{R}+\right.\right.\\
& \left.\frac{2\delta R}{\bar{R}^3}\partial_\beta \bar{R}\partial_\alpha \bar{R}-\frac{\partial_\beta\delta R\partial_\alpha \bar{R}}{\bar{R}^2}-\frac{\partial_\beta \bar{R}\partial_\alpha \delta R}{\bar{R}^2}+\bar{\Gamma}^{\kappa}_{\beta\alpha}\frac{\delta R}{\bar{R}^2}\partial_\kappa \bar{R}-\delta \Gamma^{\kappa}_{\beta\alpha}\frac{\partial_\kappa \bar{R}}{\bar{R}}-\bar{\Gamma}^{\kappa}_{\beta\alpha}\frac{\partial_\kappa\delta R }{\bar{R}}\right)\\
      & \left. -(\delta g_{\mu\nu}\bar{g}^{\alpha\beta}+\bar{g}_{\mu\nu}\delta g^{\alpha\beta})\left(\frac{\partial_\beta\partial_\alpha \bar{R}}{\bar{R}}-\frac{\partial_\beta \bar{R}\partial_\alpha \bar{R}}{\bar{R}^2}-\bar{\Gamma}^{\kappa}_{\beta\alpha}\frac{\partial_\kappa \bar{R}}{\bar{R}}\right)\right]\,,
    \end{split}
\end{equation}
with the bars denoting again background quantities. From this formula one obtains, 

\begin{equation}
    \begin{split}
        \delta T_{00}^{\rm (vac)}=&-\frac{a^2\nu\delta R}{16\pi G_N}+\frac{\nu}{8\pi G_N}\left(\frac{\partial_i\partial_i \delta R}{\bar{R}}-\frac{\bar{R}^\prime h^\prime}{2\bar{R}}+3\mathcal{H}\frac{\bar{R}^\prime}{\bar{R}^2}\delta R-3\mathcal{H}\frac{\delta R^\prime}{\bar{R}}\right)\,,\\
        \delta T_{0i}^{\rm (vac)}=&\frac{\nu}{8\pi G_N}\left(\frac{\partial_i \delta R^\prime}{\bar{R}}-\frac{\bar{R}^\prime}{\bar{R}^2}\partial_i\delta R-\frac{\mathcal{H}}{\bar{R}}\partial_i \delta R\right)\,,\\
        \delta T^{\rm (vac)}_{ij}\, =& h_{ij}\left[-a^2\left(c_0-\frac{\nu \bar{R}}{16\pi G_N}\right)+\frac{\nu}{8\pi G_N}\left(\frac{\bar{R}^{\prime\prime}}{\bar{R}}+\frac{\mathcal{H}\bar{R}^\prime}{\bar{R}}-\frac{(\bar{R}^\prime)^2}{\bar{R}^2}\right)\right]-\frac{\nu }{16\pi G_N}\frac{\bar{R}^\prime}{\bar{R}}h^\prime_{ij}+\frac{\nu}{8\pi G_N}\frac{\partial_i\partial_j\delta R}{\bar{R}}\\
&+\frac{\nu}{8\pi G_N}\delta_{ij}\left[\frac{a^2}{2}\delta R-\frac{\mathcal{H} \bar{R}^\prime\delta R}{\bar{R}^2}+\frac{\mathcal{H}}{\bar{R}}\delta R^\prime-\frac{\partial_l\partial_l\delta R}{\bar{R}}+\frac{\bar{R}^\prime h^\prime}{2\bar{R}}+\frac{\delta R^{\prime\prime}}{\bar{R}}-\frac{\bar{R}^{\prime\prime}}{\bar{R}^2}\delta R+\frac{2\delta R}{\bar{R}^3}(\bar{R}^\prime)^2-\frac{2\bar{R}^\prime\delta R^\prime}{\bar{R}^2}\right]\,.
    \end{split}
\end{equation}


\subsection{Modified Einstein equations}\label{sec:mee}

The perturbed covariant conservation equations for matter and radiation take exactly the same form as in the $\Lambda$CDM. We refer the reader to Ref. \cite{MaBertschinger} for the corresponding equations. The perturbed Einstein equations change, though. From Eq. \eqref{eq:modiEi} we obtain,

\begin{equation}\label{eq:PertEin}
\delta G_{\mu\nu}=\frac{8\pi G_N}{d+\nu\ln\left(\frac{\bar{R}}{\bar{R}_0}\right)}(\delta T^{(M)}_{\mu\nu}+\delta T^{\rm (vac)}_{\mu\nu})-\frac{\nu\,\bar{G}_{\mu\nu}}{d+\nu\ln\left(\frac{\bar{R}}{\bar{R}_0}\right)} \frac{\delta R}{\bar{R}}\,.
\end{equation}
Making use of the results presented in the preceding sections of this appendix and considering values of $|\epsilon|\ll 10^{-2}$ such that $|\epsilon|\ln(R/R_0)\ll 1$, we get, 

\begin{equation}\label{eq:pert1}
\mathcal{H}h^\prime-2\eta k^2 \, =\frac{8\pi G_N a^2\delta\rho}{d\left[1+\epsilon\ln\left(\frac{R}{R_0}\right)\right]}+\epsilon\left[\delta R\left(-\frac{a^2}{2}-\frac{k^2}{R_L}+3\mathcal{H}_L\frac{R_L^\prime}{R_L^2}-\frac{3\mathcal{H}_L^2}{R_L}\right)-\frac{h^\prime R_L^\prime}{2R_L}-3\mathcal{H}_L\frac{\delta R^\prime}{R_L}\right]\,,
\end{equation}

\begin{equation}\label{eq:etaprime}
-2\eta^\prime k^2 = \frac{-8\pi G_Na^2}{d\left[1+\epsilon\ln\left(\frac{R}{R_0}\right)\right]}(\bar{\rho}+\bar{p})\theta+\epsilon k^2\left(-\frac{\delta R^\prime}{R_L}+\frac{R_L^\prime}{R^2_L}\delta R+\frac{\mathcal{H}_L}{R_L}\delta R\right)\,,
\end{equation}

\begin{equation}
\frac{1}{2}(h^{\prime\prime}+6\eta^{\prime\prime})+\mathcal{H}(h^\prime+6\eta^\prime)-\eta k^2 =\frac{-12\pi G_Na^2}{d\left[1+\epsilon\ln\left(\frac{R}{R_0}\right)\right]}(\bar{\rho}+\bar{p})\sigma-\epsilon\left[\frac{R_L^\prime}{2R_L}(h^\prime+6\eta^\prime)+k^2\frac{\delta R}{R_L}\right]\,,
\end{equation}

\begin{equation}\label{eq:pert4}
    \begin{split}
        -h^{\prime\prime}-2\mathcal{H}h^\prime+2\eta k^2  =&\epsilon\left[\delta R\left(\frac{3}{2}a^2-3\mathcal{H}_L\frac{R^\prime_L}{R^2_L}+\frac{2k^2}{R_L}-\frac{3R^{\prime\prime}_L}{R_L^2}+\frac{6(R_L^\prime)^2}{R_L^3}+\frac{3}{R_L}(2\mathcal{H}_L^\prime+\mathcal{H}_L^2)\right)\right.\\
        &\left.+\delta R^\prime\left(\frac{3\mathcal{H}_L}{R_L}-\frac{6R^\prime_L}{R_L^2}\right)+\frac{3\delta R^{\prime\prime}}{R_L}+\frac{R^\prime_L h^{\prime}}{R_L}\right]+\frac{24\pi G_Na^2}{d\left[1+\epsilon\ln\left(\frac{R}{R_0}\right)\right]}\delta p\,.
    \end{split}
\end{equation}
The first two equations are the 00 and 0i components of Eq. \eqref{eq:PertEin}, while the last two are obtained from its ij component. They correspond to the part proportional to $\hat{k}_i\hat{k}_j$ and the trace, respectively. $\theta$ is the divergence of the 3-velocity $v^i$ in momentum space. When the various species can be treated separately, 

\begin{equation}
\delta\rho\equiv\sum_l  \delta\rho_l\,,\quad \delta p\equiv\sum_l  \delta p_l\,,\quad (\bar{\rho}+\bar{p})\theta\equiv\sum_l(\bar{\rho}_l+\bar{p}_l)\theta_l\,,\quad  (\bar{\rho}+\bar{p})\sigma\equiv\sum_l(\bar{\rho}_l+\bar{p}_l)\sigma_l\,,
\end{equation}
where the index $l$ here runs over the particle species, and the anisotropic stress is given by 

\begin{equation}
   (\bar{\rho}+\bar{p})\sigma\equiv-\left(\hat{k}_i\hat{k}_j-\frac{\delta_{ij}}{3}\right)\left(T^{i}_{j}-\delta^{i}_{j}\frac{T^{k}_{k}}{3}\right)\,. 
\end{equation}
For completeness, we also show the traceless and transverse (TT) part of the ij component of the newton constant Einstein equations, which leads to the equation for the gravitational waves,

\begin{equation}
(h^{TT}_{ij})^{\prime\prime}+\mathcal{H}(h^{TT}_{ij})^\prime\left[2+\frac{\epsilon R_L^\prime}{\mathcal{H}R_L}\right]+k^2h^{TT}_{ij}=0\,.
\end{equation} 
In StRVM gravitational waves propagate with the speed of light. Therefore, the model automatically surpasses the very tight constraints we have on this quantity thanks to the gravitational wave event GW170817 and the detection of the accompanying electromagnetic counterpart
GRB170817A \cite{Abbott2017,Creminelli2017,Ezquiaga2017}. The modification of the friction term is extremely small in all the epochs of the cosmic expansion, since it is proportional to $\epsilon$, and $|\epsilon|\lesssim 10^{-6}$. As in Horndeski theories \cite{Horndeski1974}, the correction of the friction term depends on the running of the effective Planck mass $M_*^2=1/G$ \eqref{eq:effG},

\begin{equation}\label{eq:alphaM}
\alpha_M=\frac{d\,\ln\,(M_*^2)}{d\,\ln\,a}=\frac{\epsilon R_L^\prime}{\mathcal{H}R_L}=\frac{-3\epsilon}{1+\frac{4c_0}{\rho_m}}\,.
\end{equation}
This is not surprising, since every $f(R)$ model can be reformulated as a scalar-tensor model with no kinetic term for the scalar field $\varphi$, as follows, 

\begin{equation}
 S_g=-\int d^4x \sqrt{-g}\,f(R) =- \int d^4x\sqrt{-g}\,\left[
 f(\varphi) +(R-\varphi)F(\varphi)\right]\,,
\end{equation}
with $F(R)=df/d\varphi$. This means that the action \eqref{eq:action} can be rewritten as, 

\begin{equation}\label{eq:scalrep}
S_g=-\int d^4x \sqrt{-g}\,\left[\left(c_1+c_2+c_2\ln\left(\frac{\varphi}{R_0}\right)\right)R+c_0-c_2\varphi\right]\,, 
\end{equation}
which automatically leads to \eqref{eq:alphaM}.


\subsection{Iterative method to solve the system}\label{sec:iterative}

The appearance in these equations of $\delta R$ and its first and second time derivatives, which, in turn, introduce higher derivatives of $h$ and $\eta$, complicates the implementation of these equations in \texttt{CLASS}. Nevertheless, we can apply an iterative method to solve the system of coupled differential equations formed by the Eqs. \eqref{eq:pert1}-\eqref{eq:pert4} and the perturbed conservation equations of the various matter species. First, we express all the perturbed quantities entering the equations as a perturbative expansion in $\epsilon$, 

\begin{equation}\label{eq:pertexp}
\begin{split}
 h=&h_{(0)}+\epsilon h_{(1)}+\epsilon^2h_{(3)}+...  \\
 \eta= &\eta_{(0)}+\epsilon \eta_{(1)}+\epsilon^2 \eta_{(3)}+...\\
 &\qquad \qquad ...
\end{split}
\end{equation}
where the subscripts $(0)$ denote the leading order terms in these expansions and the subscripts $(i)$ with $i\geq1$ their ith-order corrections. By substituting \eqref{eq:pertexp} in the perturbed equations we find that the leading terms must obey

\begin{equation}
\mathcal{H}h_{(0)}^\prime-2\eta_{(0)} k^2 \, =8\pi G a^2\delta\rho_{(0)}\,,
\end{equation}

\begin{equation}
-2\eta_{(0)}^\prime k^2 = -8\pi Ga^2(\bar{\rho}+\bar{p})\theta_{(0)}\,,
\end{equation}

\begin{equation}
\frac{1}{2}(h_{(0)}^{\prime\prime}+6\eta_{(0)}^{\prime\prime})+\mathcal{H}(h_{(0)}^\prime+6\eta_{(0)}^\prime)-\eta k^2 =-12\pi Ga^2(\bar{\rho}+\bar{p})\sigma_{(0)}\,,
\end{equation}

\begin{equation}
        -h_{(0)}^{\prime\prime}-2\mathcal{H}h_{(0)}^\prime+2\eta_{(0)} k^2  =24\pi Ga^2\delta p_{(0)}\,,
\end{equation}
together with the conservation equations at zeroth order. The higher order corrections ($i\geq1$) are computed as follows, 

\begin{equation}
\mathcal{H}h_{(i)}^\prime-2\eta_{(i)} k^2 \, =8\pi G_N a^2\delta\rho_{(i)}+\delta R_{(i-1)}\left(-\frac{a^2}{2}-\frac{k^2}{R_L}+3\mathcal{H}_L\frac{R_L^\prime}{R_L^2}-\frac{3\mathcal{H}_L^2}{R_L}\right)-\frac{h_{(i-1)}^\prime R_L^\prime}{2R_L}-3\mathcal{H}_L\frac{\delta R_{(i-1)}^\prime}{R_L}\,,
\end{equation}

\begin{equation}
-2\eta_{(i)}^\prime k^2 = -8\pi G a^2(\bar{\rho}+\bar{p})\theta_{(i)}+ k^2\left(-\frac{\delta R_{(i-1)}^\prime}{R_L}+\frac{R_L^\prime}{R^2_L}\delta R_{(i-1)}+\frac{\mathcal{H}_L}{R_L}\delta R_{(i-1)}\right)\,,
\end{equation}

\begin{equation}
\frac{1}{2}(h_{(i)}^{\prime\prime}+6\eta_{(i)}^{\prime\prime})+\mathcal{H}(h_{(i)}^\prime+6\eta_{(i)}^\prime)-\eta_{(i)} k^2 =-12\pi Ga^2(\bar{\rho}+\bar{p})\sigma_{(i)}-\frac{R_L^\prime}{2R_L}(h_{(i-1)}^\prime+6\eta_{(i-1)}^\prime)-k^2\frac{\delta R_{(i-1)}}{R_L}\,,
\end{equation}

\begin{equation}
    \begin{split}
        -h_{(i)}^{\prime\prime}-2\mathcal{H}h_{(i)}^\prime+2\eta_{(i)} k^2  =&\delta R_{(i-1)}\left(\frac{3}{2}a^2-3\mathcal{H}_L\frac{R^\prime_L}{R^2_L}+\frac{2k^2}{R_L}-\frac{3R^{\prime\prime}_L}{R_L^2}+\frac{6(R_L^\prime)^2}{R_L^3}+\frac{3}{R_L}(2\mathcal{H}_L^\prime+\mathcal{H}_L^2)\right)\\
        &+\delta R_{(i-1)}^\prime\left(\frac{3\mathcal{H}_L}{R_L}-\frac{6R^\prime_L}{R_L^2}\right)+\frac{3\delta R_{(i-1)}^{\prime\prime}}{R_L}+\frac{R^\prime_L h_{(i-1)}^{\prime}}{R_L}+24\pi Ga^2\delta p_{(i)}\,,
    \end{split}
\end{equation}
plus, again, the corresponding conservation equations. This is nothing more than simple perturbation theory, of course. See Sec. \ref{sec:perturbations} for an study of the convergence of this iterative method, further comments and results.


\section{Breakdown contributions to $\chi^2_{\rm min}$}\label{sec:extraTables}

In this appendix we show the additional Tables \ref{tab:table_chi2_I}-\ref{tab:table_chi2_III}, which contain the contributions of the various observables to the total $\chi^2_{\rm min}$ for all our fitting analyses.

\begin{table*}[h!]
\centering
\begin{tabular}{|c ||c | c || c| c|| c| c|    }
 \multicolumn{1}{c}{} & \multicolumn{1}{c}{} & \multicolumn{1}{c}{} & \multicolumn{1}{c}{} & \multicolumn{1}{c}{} \\
\multicolumn{1}{c}{} &  \multicolumn{2}{c}{Planck18} &  \multicolumn{2}{c}{Planck18+SNIa+BAO+CCH}&  \multicolumn{2}{c}{Planck18+SNIa\_$H_0$+BAO+CCH}
\\\hline
{\small $\chi^2_i$} & {\small $\Lambda$CDM}  & {\small StRVM}& {\small$\Lambda$CDM} & {\small StRVM}& {\small$\Lambda$CDM} & {\small StRVM}
\\\hline
{\small $\chi^2_{\rm CMB}$} & 2767.27  &  2767.28& 2768.32 & 2767.77 & 2773.19 & 2775.60
\\\hline
{\small $\chi^2_{\rm SNIa}$} & -  &-& 1458.48 & 1458.66& 1493.36 & 1471.22
\\\hline
{\small $\chi^2_{\rm BAO}$} & -  & -& 15.11 & 16.17 & 13.56 & 13.66
\\\hline
{\small $\chi^2_{\rm CCH}$} & -  &- & 13.42 & 13.25 & 13.14 & 12.70
\\\hline
{\small $\chi^2_{\rm min}$} &  2767.97 & 2767.28 &  4255.34 & 4255.30 & 4293.24 &  4273.16
\\\hline
\end{tabular}
\caption{Contribution of the $\chi^2_i$ of each observable to $\chi^2_{\rm min}$, for the $\Lambda$CDM and StRVM, and using the datasets specified in the upper part of the table. The corresponding fit values of the parameters are reported in Table \ref{tab:tableI}.}
\label{tab:table_chi2_I}
\end{table*}

\begin{table*}[h!]
\centering
\begin{tabular}{|c ||c | c || c| c|| c| c|    }
 \multicolumn{1}{c}{} & \multicolumn{1}{c}{} & \multicolumn{1}{c}{} & \multicolumn{1}{c}{} & \multicolumn{1}{c}{} \\
\multicolumn{1}{c}{} &  \multicolumn{2}{c}{Planck18(np)} &  \multicolumn{2}{c}{Planck18(np)+SNIa+BAO+CCH}&  \multicolumn{2}{c}{Planck18(np)+SNIa\_$H_0$+BAO+CCH}
\\\hline
{\small $\chi^2_i$} & {\small $\Lambda$CDM}  & {\small StRVM}& {\small$\Lambda$CDM} & {\small StRVM}& {\small$\Lambda$CDM} & {\small StRVM}
\\\hline
{\small $\chi^2_{\rm CMB}$} &  1179.54 & 1178.80 & 1181.31  & 1181.24 &  1183.00 & 1184.69
\\\hline
{\small $\chi^2_{\rm SNIa}$} & -  & -& 1459.11 & 1459.21 & 1494.81 & 1466.41
\\\hline
{\small $\chi^2_{\rm BAO}$} &  - & - & 13.89  & 14.17 & 13.42 & 13.38
\\\hline
{\small $\chi^2_{\rm CCH}$} & -  & - & 13.45 & 12.74 & 13.20 & 13.37
\\\hline
{\small $\chi^2_{\rm min}$} &  1179.54 & 1178.80 & 2667.78 & 2667.34 & 2704.42 & 2677.84 
\\\hline
\end{tabular}
\caption{Same as in Table \ref{tab:table_chi2_I}, but for the datasets that do not incorporate the high-$\ell$ CMB polarization data from {\it Planck}. The corresponding fit values of the parameters are reported in Table \ref{tab:tableIII}.}
\label{tab:table_chi2_II}
\end{table*}

 \begin{table*}[h!]
\centering
\begin{tabular}{|c ||c | c || c| c|| c| c|    }
 \multicolumn{1}{c}{} & \multicolumn{1}{c}{} & \multicolumn{1}{c}{} & \multicolumn{1}{c}{} & \multicolumn{1}{c}{} \\
\multicolumn{1}{c}{} &  \multicolumn{2}{c}{Planck18(lens)} &  \multicolumn{2}{c}{Planck18(lens)+SNIa+BAO+CCH}&  \multicolumn{2}{c}{Planck18(lens)+SNIa\_$H_0$+BAO+CCH}
\\\hline
{\small $\chi^2_i$} & {\small $\Lambda$CDM}  & {\small StRVM}& {\small$\Lambda$CDM} & {\small StRVM}& {\small$\Lambda$CDM} & {\small StRVM}
\\\hline
{\small $\chi^2_{\rm CMB}$} &   &  & 2778.64 & 2777.07 & 2780.42  & 2785.70 
\\\hline
{\small $\chi^2_{\rm SNIa}$} & -  & -& 1458.26 & 1458.43 &  1496.08 & 1471.12
\\\hline
{\small $\chi^2_{\rm BAO}$} &  - & - & 15.91  & 16.27 & 13.59 & 13.88
\\\hline
{\small $\chi^2_{\rm CCH}$} & -  & - & 13.44 & 13.18 & 13.26 & 12.71 
\\\hline
{\small $\chi^2_{\rm min}$} & 2775.28 & 2775.22 &  4266.26 & 4264.94 & 4303.35 & 4283.43 
\\\hline
\end{tabular}
\caption{Same as in Table \ref{tab:table_chi2_I}, but including the CMB lensing information from {\it Planck}. The corresponding fit values of the parameters are reported in Table \ref{tab:tablelens}.}
\label{tab:table_chi2_III}
\end{table*}


 \end{document}